\documentclass[%
aps,
prd,
twocolumn,
nofootinbib,
amsmath,amssymb,
aps,
]{revtex4-2}
\usepackage{graphicx}
\usepackage{adjustbox}
\usepackage{dcolumn}
\usepackage{bm}
\usepackage[dvipsnames]{xcolor}
\usepackage[utf8]{inputenc}
\usepackage[english]{babel}
\usepackage{amsfonts}
\usepackage{amsmath}
\usepackage{amssymb}
\usepackage{import}
\usepackage{cancel}
\usepackage{feynmp-auto}
\usepackage{slashed}
\usepackage{mathrsfs}
\usepackage{float}
\usepackage{verbatim}
\usepackage{color, colortbl} 
\usepackage{upgreek}
\usepackage{comment}
\usepackage{multirow}
\usepackage{enumitem}
\usepackage{mathtools}
\usepackage[vcentermath]{youngtab}

\usepackage[T1]{fontenc} 

\usepackage{array}
\usepackage{pifont} 
\usepackage{ytableau} 
\usepackage{hyperref}
\hypersetup{
    colorlinks=true,
    menucolor = teal,
    linkcolor= teal,
    filecolor=teal,      
    urlcolor= teal,
    citecolor = teal, 
    pdftitle={Overleaf Example},
    pdfpagemode=FullScreen,
    }
\def\sv#1{{\bf 
\textcolor{blue}{[SV: {#1}]}}}

\begin{document}

\preprint{ULB-TH/26-01}

\title{Constraining long-lived dark sector particles with CMB and Lyman-$\alpha$}

\author{Laura Lopez-Honorez$^{a, b}$}
\author{Sonali Verma$^{a}$}%
\affiliation{${}^a$Service de Physique Th\'eorique, \href{https://ror.org/01r9htc13}{ Universit\'e Libre de Bruxelles}, B-1050 Brussels, Belgium}
\affiliation{${}^b$Theoretische Natuurkunde \& The International Solvay Institutes, \href{https://ror.org/006e5kg04}{Vrije Universiteit Brussel}, B-1050 Brussels, Belgium}


\begin{abstract}
We use measurements of the intergalactic medium (IGM) temperature from the Lyman-$\alpha$ forest to place new limits on models in which long-lived dark sector (DS) particles, with lifetimes longer than $10^{16}$ s, deposit energy into the IGM through their decays.  Such DS decays into Standard Model (SM) states can modify the late-time thermal history of the IGM, making Lyman-$\alpha$ data a sensitive probe of hidden sectors with cosmologically long lifetimes. Our analysis demonstrates that constraints from late-time IGM heating offer a complementary window to those from the Cosmic Microwave Background (CMB), in constraining dark sector parameter space. We further revisit limits on such decaying DS models from Planck's measurements of the optical depth to reionization and provide updates relevant for DS lifetimes longer than $10^{14}$ s. The model-independent constraints on the DS parameter space we derive in this work can be reinterpreted for a wide range of decaying hidden-sector scenarios, including evaporating primordial black holes and SM-coupled dark photons.
\end{abstract}

\maketitle
\section{Introduction}


Interactions in the dark sector such as dark matter (DM) annihilations and decays to Standard Model (SM) particles can lead to exotic energy injection, consequently producing significant modifications in the ionization and thermal history of the early Universe, see
e.g.~\cite{Shull:1985, Adams:1998nr,  Chen:2003gz, Padmanabhan:2005es,Slatyer:2009yq, Slatyer:2015jla, Slatyer:2015kla, Slatyer:2016qyl,Poulin:2016anj, Lopez-Honorez:2013cua,Diamanti:2013bia, Lopez-Honorez:2016sur,  Liu:2016cnk,Slatyer:2016qyl,   Acharya:2019uba, Lucca:2019rxf,Bolliet:2020ofj,  Capozzi:2023xie, Liu:2023nct, Myers:2025pfx}.  Constraints on the free electron fraction from measurements of the Cosmic Microwave Background (CMB) anisotropy spectra by \emph{Planck} data have been widely used to set stringent limits on the dark matter annihilation cross section~\cite{Planck:2018vyg} and the DM decay lifetime (see Refs.~\cite{Adams:1998nr, Acharya:2019uba, Poulin:2016anj, Lucca:2019rxf, Capozzi:2023xie, Liu:2023nct}).

Late-time probes of the intergalactic medium (IGM), such as IGM temperature measurements from the Lyman-$\alpha$ (Ly$\alpha$) forest of quasars (QSOs) (see e.g.~\cite{Walther:2018pnn, Gaikwad:2020art}) in the redshift range $z \lesssim 5$, can constrain new dark interactions that modify the IGM temperature. These measurements have already been used to set limits on DM annihilations~\cite{Cirelli:2009bb} and DM decay lifetimes~\cite{Diamanti:2013bia, Liu:2016cnk, Liu:2020wqz,Capozzi:2023xie}. The derived Ly$\alpha$ constraints in Ref.~\cite{Liu:2020wqz} in particular were demonstrated to be even stronger than CMB limits for DM decaying to electron--positron pairs, with lifetime longer than the age of the Universe, in the $1-10~\mathrm{MeV}$ mass range~\cite{Liu:2020wqz}.

Upcoming measurements of the 21-cm power spectrum from ongoing experiments such as HERA~\cite{DeBoer:2016tnn}, SKA\cite{inproceedings}, will further probe another redshift range, $z \sim 5$--$20$. Limits on DM lifetime from prospective 21-cm power spectrum measurements in Refs.~\cite{Facchinetti:2023slb, Sun:2023acy, Agius:2025nfz} show their potential to provide the most stringent limits on the DM lifetime in the mass range below $\sim 2~\mathrm{GeV}$.

In addition to exotic energy injection by dark matter interactions, other metastable dark sector (DS) particles, with lifetimes much shorter than the age of the Universe, can also lead to IGM modifications through their decay to SM particles. Such particles arise in many beyond-Standard-Model (BSM) extensions: axion-like particles (ALPs) coupling to SM photons and electrons, (which do not constitute DM), (see Ref.~\cite{Capozzi:2023xie, Langhoff:2022bij} for limits on DM ALPs and non-DM ALPs respectively), massive dark photons coupling to the SM electromagnetic sector, light primordial black holes (PBHs) that evaporate to SM particles via Hawking radiation~\cite{Poulin:2016anj, Khan:2025kag, 
Sun:2025ksr, Saha:2024ies}, and composite dark sectors with long-lived resonances, among others.

Cosmological constraints on these models become crucial for small couplings, corresponding to DS lifetimes $\tau_{\rm DS} \gg 1~\mathrm{s}$, that are not probed at terrestrial experiments. While constraints from Big Bang nucleosynthesis (BBN) and CMB spectral distortions are sensitive to the range $\tau_{\rm DS} \sim 10^{4}$ -- $10^{10}~\mathrm{s}$~\cite{Poulin:2016anj, Lucca:2019rxf}, probes such as the CMB anisotropies, Ly$\alpha$, and 21-cm become more relevant for much longer DS lifetimes.  CMB is most sensitive to DS lifetimes $\tau_{\rm DS} \gtrsim H^{-1}(z_{\rm rec})\sim10^{14}~\mathrm{s}$, where $H(z)$ denotes the Hubble rate and $z_{\rm rec}\sim 1000$ is the recombination redshift, and late-time probes such as the Lyman$-\alpha$ probe longer DS lifetimes, $\tau_{\rm DS}  \gtrsim H^{-1}(z=5)\sim 10^{17}~\mathrm{s}$.

In light of this, bounds have been derived on many such DS models using current cosmological probes in Refs.~\cite{Poulin:2016anj, Lucca:2019rxf, Khan:2025kag, Saha:2024ies, Caputo:2025avc, Langhoff:2022bij, Acharya:2019uba}. References~\cite{Caputo:2025avc, Langhoff:2022bij} have reinterpreted DM constraints to set CMB limits on dark photon and ALP parameter space, respectively. Recent works~\cite{Saha:2024ies, Khan:2025kag} have set limits on PBH parameter space using Ly$\alpha$ heating, whereas Refs.~\cite{Poulin:2016anj, Slatyer:2016qyl, Acharya:2019uba, Slatyer:2012yq} and Ref.~\cite{Lucca:2019rxf} set model-independent limits from the CMB \textit{Planck 2015} data~\cite{Planck:2015fie} and \textit{Planck 2018} data~\cite{Planck:2018vyg} respectively on DS parameter space spanned by the DS lifetime $\tau_{\rm DS}$ and the fraction of the DM density in DS particles, $f_{\rm DS}$. 

 In this work, we set new limits on the DS parameter space from existing IGM temperature measurements from Ly$\alpha$ spectra~\cite{Walther:2018pnn, Gaikwad:2020art}.
While our main goal is to constrain DS using Ly$\alpha$, we also revisit bounds on DS parameter space from the CMB (see Refs.~\cite{Poulin:2016anj, Slatyer:2016qyl}) using the latest \emph{Planck 2018} measurement of the optical depth of reionization~\cite{Planck:2018vyg}. We note that Ref.~\cite{Lucca:2019rxf} has already presented \emph{Planck 2018} constraints on the DS parameter space. However, the authors of \cite{Lucca:2019rxf}, in performing their analysis relied under a simplifying assumption that all the energy emitted by the decaying DS is instantaneously deposited in the IGM with no energy lost due to secondary interactions. In their notation, this corresponds to setting the so-called efficiency function to a constant $f_{\rm eff} = 1$. The fraction of deposited energy channelled into hydrogen ionization is then given by $f_{\rm H \ ion} = f_{\rm eff} \chi_{\rm ion}$ where $\chi_{\rm ion}$ varies between $0.001-0.35$.~\footnote{Ref.~\cite{Lucca:2019rxf} used pre-tabulated values of $\chi_{\rm ion}$, the partition fraction for the hydrogen ionization channel from Ref.~\cite{Galli:2013dna}, which depends only on $x_e$, the free electron fraction. This quantity varies from \(10^{-3}\) to \(0.35\) as \(x_e\) decreases from \(x_e \simeq 1\) to \(x_e \simeq 10^{-4}\)~\cite{Galli:2013dna}.}

As we show in Sec.~\ref{sec:IGM_history} this assumption  is a a priori optimistic. Our energy deposition functions denoted as $f_{c}(z)$, are typically much smaller than unity in all deposition channels. This motivates our re-analysis of \emph{Planck 2018} bounds.  In this work, we carefully compute the efficiency of energy deposition, focusing on DS decaying to $e^+e^-$ and $\gamma \gamma$. This allows us to consistently compare our Ly$\alpha$ limits against those obtained from the latest CMB measurements.  

Using IGM temperature measurements from Ly$\alpha$ spectra, our analysis shows that bounds from Ly$\alpha$ can be competitive with previous CMB bounds from \textit{Planck 2015} for long DS lifetimes close to $10^{17}$ s. Our analysis adopts a conservative approach in which we do not add any astrophysical sources of photoheating. Including such sources of photoheating  have been shown to tighten limits $\sim 2-8$ times for the scenarios considered in~\cite{Liu:2020wqz}, demonstrating that even under conservative assumptions, Ly$\alpha$ measurements provide a complementary probe for late-time energy injections from DS decays. \\
In deriving our own CMB bounds, we further employ the same deposition efficiency functions used in the Ly$\alpha$ analysis, enabling a fully consistent comparison. This is important because updates in methodologies for computing energy deposition can lead to non-negligible differences when comparing bounds derived across different works. Compared to previous studies~\cite{Lucca:2019rxf, Poulin:2016anj}, we have additionally taken account of the \emph{backreaction effect} in computing the efficiency functions, which have been shown to impact IGM observables in Refs.~\cite{Liu:2020wqz, Liu:2019bbm}. \\
We also treat DS decays into electron-positron $(e^+e^-)$ and photons ($\gamma$) separately, in contrast to the \emph{Planck 2018} CMB analysis in \cite{Lucca:2019rxf}. This allows us to better understand the impact of varying the DS particle mass or equivalently the injected energy, on the resulting bounds for the two decay channels. In particular, we find that for the DS decay to $e^+e^-$ case, the tightest bounds from CMB are obtained for a DS mass of 100 MeV, whereas for DS decay to $\gamma \gamma$, the tightest bounds are obtained for DS mass close to 40 eV. While Ref.~\cite{Poulin:2016anj} considered both $e^+e^-$ and $\gamma \gamma$ injection channels in deriving CMB bounds using a full Monte Carlo analysis of the \textit{Planck 2015} data, their  study was restricted to injection energies above 10 keV. In contrast, our work extends to lower injection energies (DS mass) as low as 40 eV for the case of DS decay to photons. \\ Finally, as an application of our analysis, we translate our bounds into exclusions for evaporating PBHs, demonstrating that bounds from Lyman-$\alpha$ can be competitive for PBH masses larger than $\sim 3 \times 10^{14}$ g (see also recent works~\cite{Saha:2024ies, Khan:2025kag}).

To build intuition for the results presented in this work, we now estimate how energy injection from DS decay modifies the thermal and ionization history of the IGM. Let us consider a generic DS particle with mass $m_{\rm DS}$, comprising a fraction $f_{\rm DS} = \Omega_{\rm DS}/\Omega_{\rm DM}$ of the total dark matter density, decaying into SM particles with a lifetime $\tau_{\rm DS}$. Then the energy injection rate due to its decay is given by 
\begin{align}\label{eq:inj_DS}
    \bigg( \frac{dE}{dV dt} \bigg)^{\rm DS, \ inj} = f_{\rm DS} \ e^{-t/\tau_{\rm  DS}} \rho_{\rm DM}^{0} \ (1 + z)^3 \frac{1}{\tau_{\rm DS}},
\end{align}
where $\tau_{\rm DS} < t_{\rm uni}$ with $t_{\rm uni}$ being the age of the Universe, and $\rho_{\rm DM, 0}$ is the DM density today~\footnote{For a stable DM particle (with $f_{\rm DS} = 1)$, $\tau \gg t_{\rm uni}$, $e^{-t/\tau_{\rm  DS}} \approx 1$.}.
This injected energy in turn can leave an imprint on cosmological probes by modifying the free electron fraction $x_{e}$, and/or IGM temperature $T_{\rm m}$. The modification in the two observables is roughly given by the energy injected by the DS decay per baryon~\footnote{Note that in this estimate, we naively assume that all the energy injected from the DS decay goes into ionizing/heating the IGM.} (see, e.g.~Ref.~\cite{Slatyer22,GGI:2025Lectures, Palomares26})
\begin{equation}\label{eq:deltaTm}
    \Delta T_{\rm m} \simeq \frac{1}{n_b}\bigg( \frac{dE}{dV dz} \bigg)^{\rm DS, \ inj}, \  \Delta x_{e} \simeq \frac{1}{n_b} \frac{1}{E_{\rm ion}} \bigg( \frac{dE}{dV dz} \bigg)^{\rm DS, \ inj},
\end{equation}
where the injected energy is defined in Eq.~\eqref{eq:inj_DS}, and can be translated to the form used in Eq.~\eqref{eq:deltaTm} using $dt/dz = (H(z)(1+z))^{-1}$. For $\Delta x_e$, we have normalized the injected energy by the hydrogen ionization energy,~$E_{\rm ion}$. 

We can now estimate the change in $x_e$ from DS decay as
\begin{align}
    \label{eq:stima_xe}
    \Delta x_{\rm e} &\simeq \frac{1}{n_b} \frac{1}{E_{\rm ion}} f_{\rm DS} \frac{e^{-t/\tau_{\rm DS}}}{\tau_{\rm DS}} \rho_{\rm DM}^0\frac{(1 + z)^2}{H(z)}\,. 
\end{align}
We can use the scaling for the baryon number density with redshift, $n_b = n_b^0(1 + z)^3$ where $n_b^0 = \rho_b^0/m_b$ with baryon mass $m_b \sim 1 \ \mathrm{GeV}$ and approximate   $E_{\rm ion} \sim 10$ eV. Furthermore, setting $z = 1000$ with the Hubble parameter $H(z \sim 1000) = 10^{-14} \ \mathrm{s}^{-1}$, $e^{-t(z\sim1000)/\tau_{\rm DS}} \simeq 1,$ we get
\begin{equation}
 \label{eq:stima_xez1000}
  \Delta x_{\rm e}(z=1000)    \sim 10^{-3}  \  \bigg(\frac{f_{\rm DS}}{10^{-9}} \bigg) \bigg(\frac{ \rho_{\rm DM}^0/\rho_{\rm b}^0}{5} \bigg) \bigg( \frac{5 \times 10^{13} \ \mathrm{s}}{\tau_{\rm DS}} \bigg)  \nonumber .
\end{equation}
This estimate shows that a DS decay with the above parameters can modify $x_e$ at the $\mathcal{O}(10^{-3})$ level, enough to leave an observable imprint on the CMB. This estimate can be compared to current constraints on DS parameter space derived using CMB $\textit{Planck 2015}$ data in Ref.~\cite{Slatyer:2016qyl} which exclude $f_{\rm DS} \gtrsim 10^{-10} - 10^{-11}$ for DS lifetime $\tau_{\rm DS} \sim 10^{13}$s.

Next, we want to understand how the IGM temperature will be modified by such decaying DS particles. Using Eqs.~\eqref{eq:inj_DS} and~\eqref{eq:deltaTm}, we can now approximate the change in the IGM temperature as
\begin{align}\label{eq:stima_Tm}
    \Delta T_{\rm m} &\simeq \frac{1}{n_b} f_{\rm DS} \frac{e^{-t/\tau_{\rm DS}}}{\tau_{\rm DS}} \rho_{\rm DM}^0\frac{(1 + z)^2}{H(z)}\,. 
\end{align}
 Considering $z=5$ with $H(z \sim 5) = 10^{-17} \ \mathrm{s}^{-1}$, we get 
 \begin{equation}
 \label{eq:stima_Tmz5}
      \Delta T_{\rm m}(z=5)\sim 10^4 \ \mathrm{K} \  \bigg(\frac{f_{\rm DS}}{10^{-9}} \bigg) \bigg(\frac{ \rho_{\rm DM}^0/\rho_{\rm b}^0}{5} \bigg) \bigg( \frac{6 \times 10^{16} \ \mathrm{s}}{\tau_{\rm DS}} \bigg)  \ .\nonumber
 \end{equation}
A comparison of this naive estimate  with the IGM temperature measurement from Ly$\alpha$ at $T_{\rm m}(z = 5) \sim 10^4 $ K~\cite{Walther:2018pnn} already shows that Ly$\alpha$ can probe DS energy density fractions of $f_{\rm DS} \sim 10^{-9}$ for lifetimes $\tau_{\rm DS} \simeq 10^{17}$ s. In the rest of the paper, we will derive this bound using a more rigorous analysis based on the evolution of IGM ionization and thermal history, utilizing the IGM Ly$\alpha$ measurements in Gaikwad et al.~\cite{Gaikwad:2020art}, and Walther et al.~\cite{Walther:2018pnn}. For this analysis, we  use the publicly available code \texttt{DarkHistory}~\cite{Liu:2019bbm,Sun:2022djj}, which we modified to include DS energy injection. 


Beyond these existing probes, forthcoming 21-cm power spectrum measurements have the potential to improve over current CMB and Ly$\alpha$ constraints for DS interactions~\cite{Poulin:2016anj} (see also Ref.~\cite{Sun:2025ksr} for a recent analysis for PBH evaporation). Using the same back of the envelope estimate in Eq.~\eqref{eq:stima_Tm}, we can understand if this is the case for generic decaying dark sectors. Setting $\Delta T_{\rm m}/T_{\rm m} \sim \mathcal{O}(1)$, in order to have a measurable imprint, and further using that future measurements from experiments such as HERA~\cite{HERA:2021bsv} will probe $T_{\rm m}(z =15) \sim 10 $ K, we can see that future 21-cm probes can improve the DS limits to $f_{\rm DS} \sim 10^{-11}$ for DS lifetime $\tau_{\rm DS} \sim 10^{16}$ s. A more rigorous analysis based on forthcoming HERA measurement of the 21-cm power spectrum is foreseen in a future work. The above estimates are projected in our summary plot of Fig.~\ref{fig:all_cons} as dot dashed lines, see also e.g.~\cite{Cima:2025zmc} for 21-cm projections~\footnote{For the decaying DM scenario, these estimates constrain DM lifetimes of around $\tau_{\rm DM} \gtrsim 10^{28}$ s, which are in the same ballpark as the constraints derived in ~\cite{Facchinetti:2023slb}, see also discussion regarding astrophysical uncertainties in the context of 21-cm signal in Refs.~\cite{Lopez-Honorez:2013cua,Agius:2025xbj}.}.

The paper is organised as follows: in section~\ref{sec:IGM_history}, we give a summary on IGM history evolution and DS decay impact on it. In section~\ref{sec:method}, we explain the method used for our analysis and give our final constraints in the DS parameter space, comparing our obtained results with past works.  In appendices~\ref{app:cmb_mds} and~\ref{app:lyman_mds}, we discuss the impact of changing DS mass (or injected energy) on the results we obtain. Finally, in appendix~\ref{app:pbh_cons}, as an application of our analysis, we further translate our bounds to exclusions for the case of evaporating PBHs. Finally, a summary of our work can be found in  section~\ref{sec:summary}.   The codes that were used to develop this analysis will be further made public after the publication of this paper.

\begin{figure*}
     \centering
     \includegraphics[width=5.7cm]{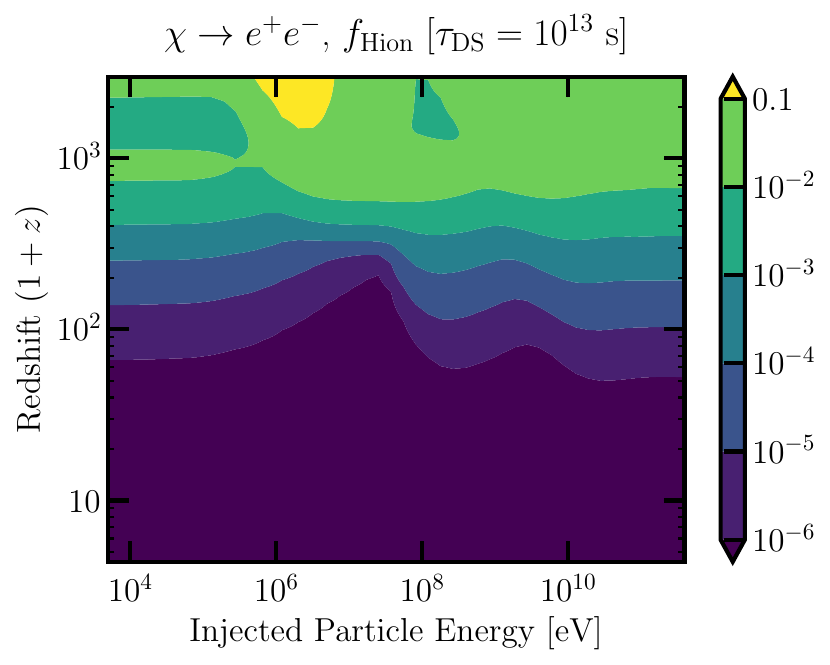}
     \hfill
     \includegraphics[width=5.7cm]
     {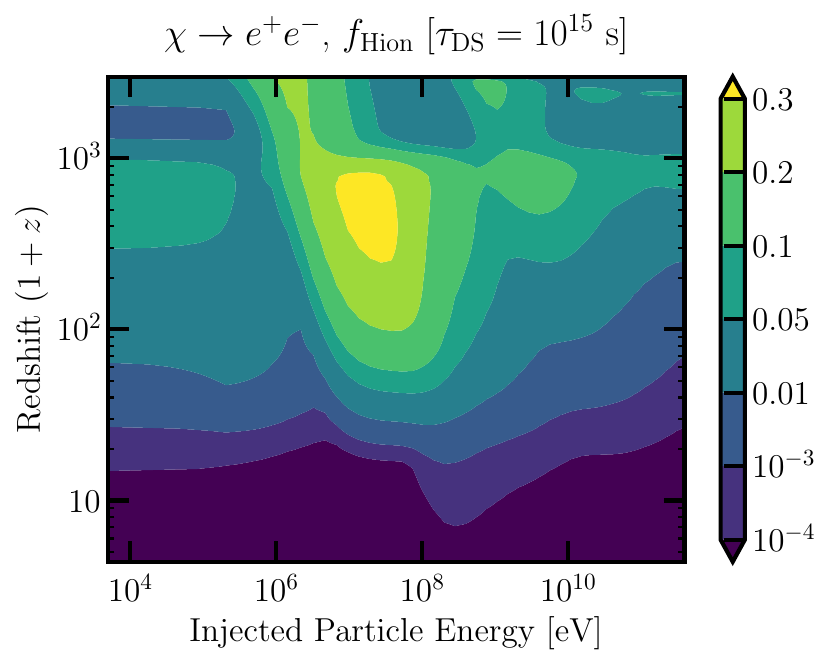}
     \hfill
     \includegraphics[width=5.7cm]{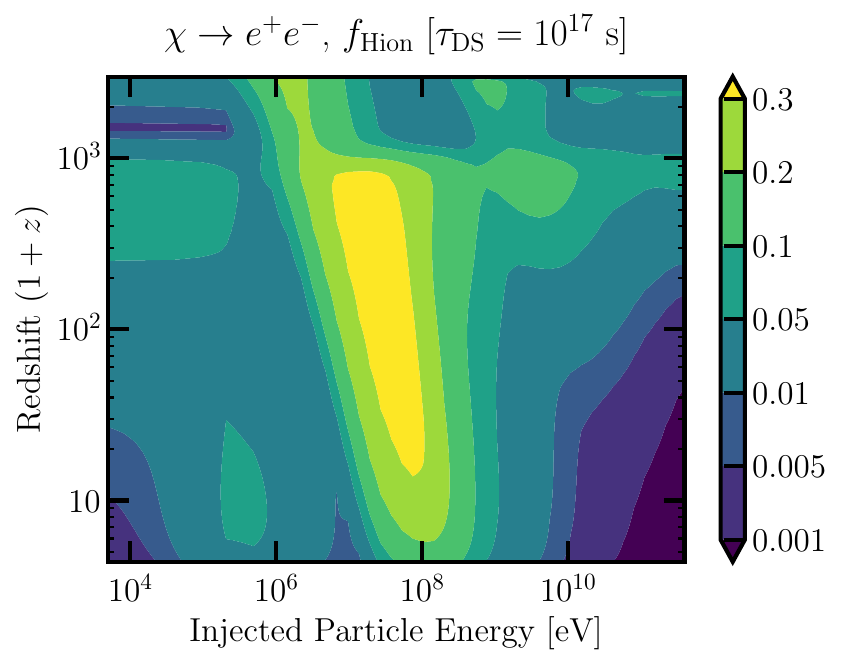}
     \hfill
     \includegraphics[width=5.7cm]{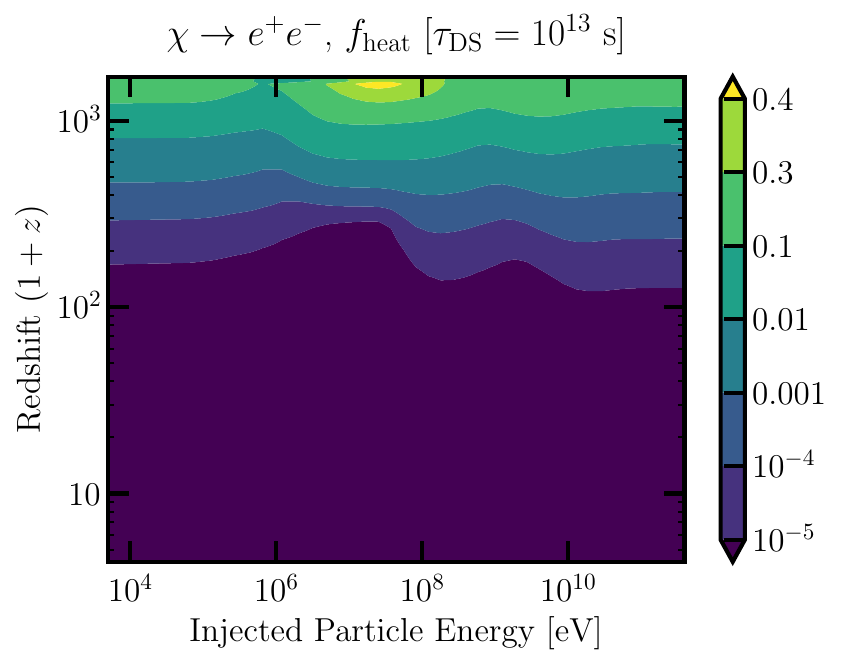}
     \hfill
     \includegraphics[width=5.7cm]
     {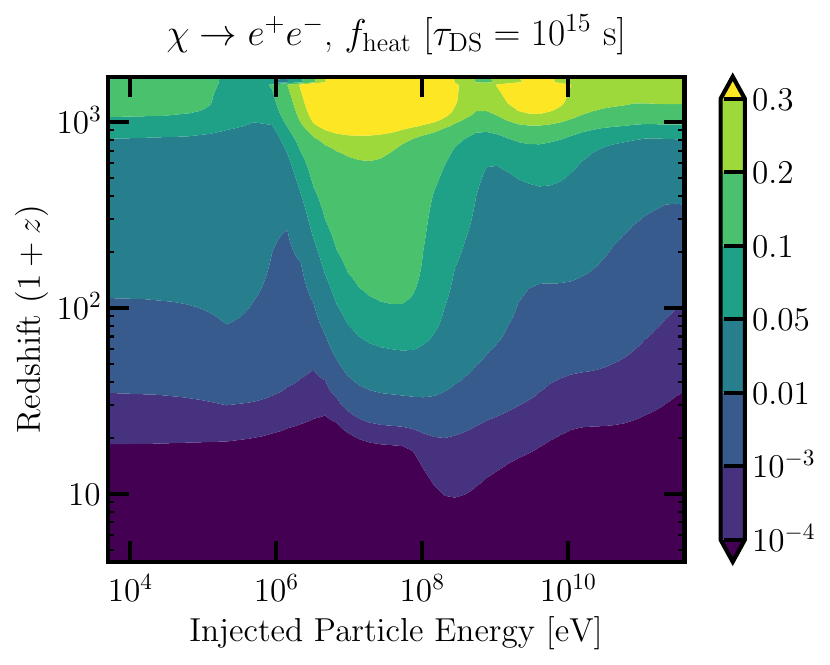}
     \hfill
     \includegraphics[width=5.7cm]{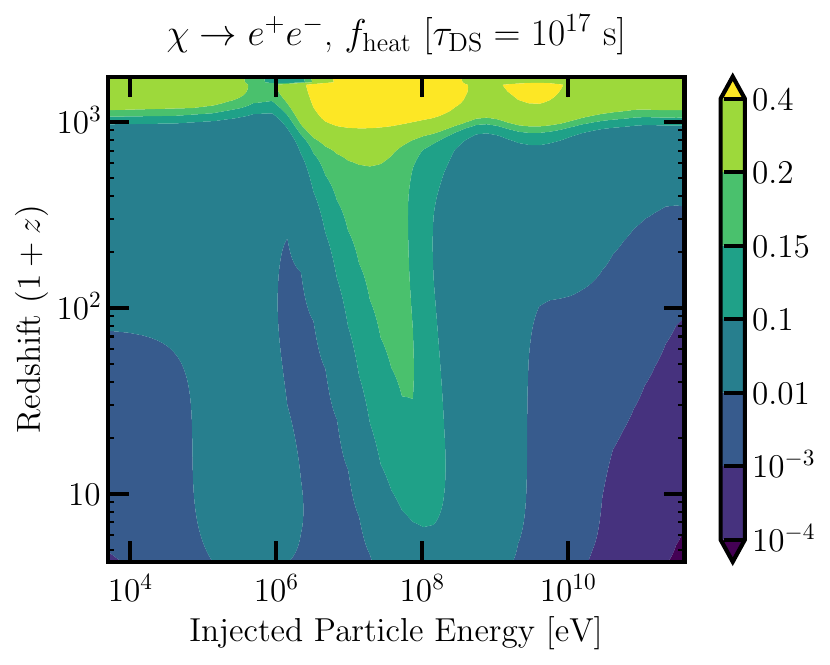}
     \caption{Computed $f_{\rm Hion}(z)$ (\textit{top row})  and $f_{\rm heat}(z)$ (\textit{bottom row}) as a function of redshift of deposition and initial kinetic energy of the injected particles, without backreaction, obtained using a modified version of \texttt{DarkHistory} code for DS particle $\chi$ decaying into $e^+e^-$ for DS lifetime $\tau_{\rm DS} = 10^{13}$ s \emph{(first column)}, $10^{15}$ s \emph{(second column)} and $10^{17}$ s \emph{(third column)}.}
    \label{fig:fc_two_column}
\end{figure*}
\section{Impact of dark sector decay on ionization and thermal history}
\label{sec:IGM_history}

In this section, we give a summary of the equations governing evolution of the IGM temperature $T_{\rm m}$ and ionized electron fraction $x_e$ following Refs.~\cite{Capozzi:2023xie, Liu:2020wqz,Liu:2019bbm} and the impact of long-lived dark sector decays on these. Such a formalism for the case of energy injection from dark matter annihilations and decays has been implemented in the publicly available code \texttt{DarkHistory}~\cite{Liu:2019bbm,Sun:2022djj} which we employ (with modifications to implement the dark sector decay case) for our analysis in Sec.~\ref{sec:method}.

The coupled system of equations tracking $T_{\rm m}$ and the different contributions to $x_e$ from hydrogen and helium ionization can be written in a compact form as~\cite{Capozzi:2023xie}

\begin{equation}\label{eq:evol_eq}
\dot{Y} = \dot{Y}^{(0)}  + \dot{Y}^{\mathrm{astro}} + \dot{Y}^{\mathrm{DS}}\, 
\end{equation}
where 
\begin{equation}
Y =
\begin{pmatrix}
T_{\rm m} \\
x_{\mathrm{HII}} \\
x_{\mathrm{HeII}} \\
x_{\mathrm{HeIII}}
\end{pmatrix}.
\end{equation}
Here the ionized fractions $x_{X}$ correspond to $x_{X} = n_{X}/n_{\rm H}$ where $n_{\rm H}$ is the number density of both neutral and ionized hydrogen. HII, HeII and HeIII stand for ionized hydrogen (H$^+$), singly ionized helium (He$^+$) and doubly ionized helium (He$^{++}$), respectively. Here, $\dot{Y}^{(0)}$ includes the terms in the coupled equations for $T_{\rm m}$ and $x_X$ in the absence of both astrophysical sources of reionization as well as exotic energy injection from dark sectors. \\
In the absence of helium evolution, the matter temperature evolution term $\dot{T}_{\rm m}^{(0)}$ includes contributions from i) adiabatic cooling, ii) Compton scattering from CMB, iii) heating/cooling due to atomic processes (recombination, collisional ionization, collisional excitation, and bremsstrahlung) (see details on each process in Ref.~\cite{Liu:2019bbm} and refs. within), whereas the ionized fraction evolution term, $\dot{x}_{\rm HII}^{(0)}$  includes contributions from atomic processes - recombination and collisional ionization. \\
The second term in Eq.~\eqref{eq:evol_eq} given by $\dot{Y}_{\rm astro}$, relevant at $z \lesssim 30$, includes photoionization and photoheating from astrophysical sources of reionization.\\
Finally, the third term in Eq.~\eqref{eq:evol_eq}, $\dot{Y}_{\rm DS}$, contains contribution from energy injection in the IGM from DS decays.  More details on the first and second terms of Eq.~\eqref{eq:evol_eq} can be found in Refs.~\cite{Liu:2019bbm, Capozzi:2023xie}, and we will now focus on describing the physics behind the DS term given by $\dot{Y}_{\rm DS}$.\\[0.1cm]

\noindent \textbf{DS decay.} The injected energy rate due to the decay of a DS particle $\chi$, given in Eq.~\eqref{eq:inj_DS}, contributes to the IGM ionization and thermal history. \\ 
The primary injected particles eventually give rise to photons, electrons and positrons which can deposit their energy into the IGM. This energy is deposited into multiple channels denoted by $c$. These are hydrogen ionization (`H ion'), helium ionization (`He ion'), hydrogen Ly$\alpha$ excitation (`exc'), heating of the IGM (`heat'), and sub-10.2 eV continuum photons (`cont')~\cite{Liu:2019bbm}. The actual deposited energy from the DS decay in each such channel ($c$) can be parametrized as~\cite{Poulin:2016anj, Slatyer:2012yq}
\begin{equation}\label{eq:fc_eqn}
    \bigg( \frac{dE}{dV dt} \bigg)^{\rm DS, \ dep}_{c} = f_c(z, \mathbf{x}) \ \bigg( \frac{dE}{dV dt} \bigg)^{\rm \rm long-lived, \ inj}
\end{equation}
where  $\big(\frac{dE}{dV dt} \big)^{\rm long-lived, \ inj}$ (following the notation of Ref.~\cite{Poulin:2016anj}) corresponds to the energy injected in the case of a long-lived DS particle with $\tau_{\rm DS} \gg t_{\rm uni}$.  Here, $f_c(z, \mathbf{x})$ is the redshift-dependent energy efficiency function for each channel, depending on the ionized fractions $\mathbf{x} = (x_{\rm HII}, x_{\rm HeII}, x_{\rm HeIII})$ at each $z$. Following ref.~\cite{Slatyer:2016qyl}, note that we have normalized the efficiency function in Eq.~\eqref{eq:fc_eqn} using the energy that would be injected without the exponential suppression, since the energy deposition rate can remain appreciable even after the energy injection has stopped, for the case of short-lifetime decays. Using this framework, we have computed $f_c(z, \mathbf{x})$ for DS decays to $e^+e^-$ and $\gamma \gamma$ final states by implementing equation Eq.~\eqref{eq:fc_eqn} in our modified version of \texttt{DarkHistory}. The results we obtain consistently reproduce the efficiency functions for short-lifetime decaying particles found in the literature (see, e.g., Fig.~1 of Ref.~\cite{Slatyer:2016qyl}), thereby providing a cross-check for our computation. We show the $f_c(z)$ functions obtained in our analysis in Fig.~\ref{fig:fc_two_column} for $c =$ H ionization (`H ion') (first row) and for $c =$ IGM heating (`heat'), for DS decaying to $e^+e^-$ pairs with DS lifetime $\tau_{\rm DS} =$  $10^{13}$ s (first column) s, $\tau_{\rm DS} =$  $10^{15}$ s (second column) and $\tau_{\rm DS} =10^{17}$ s (third column), as a function of injection energy and redshift of absorption. The former is given by the kinetic energy of one of the decay products $K_{\rm inj} = (m_{\rm DS} - 2 m_{\rm e})/2$.
\begin{figure*}
    \includegraphics[width=8.5cm]{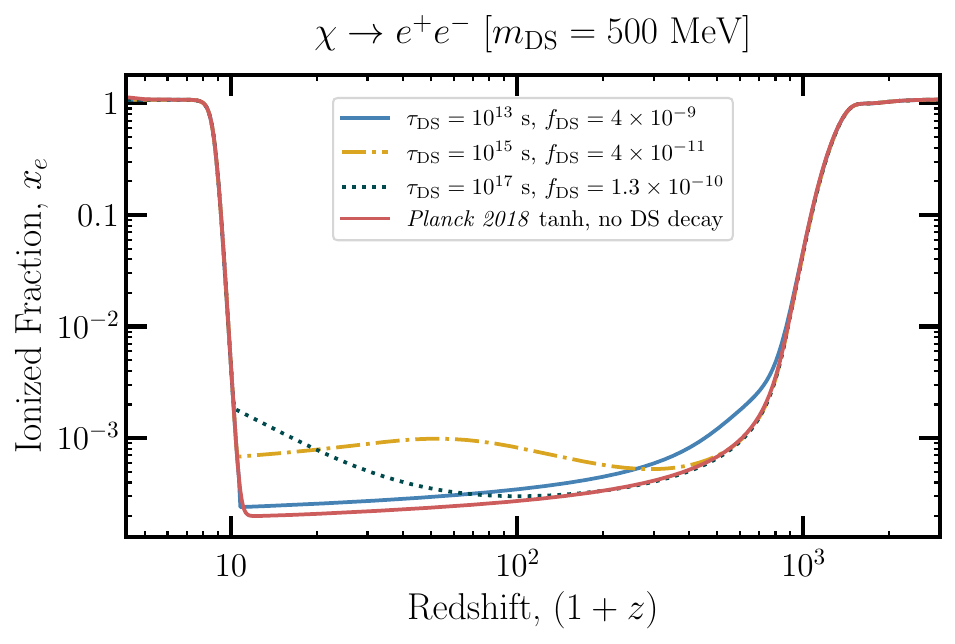}
    \hfill
    \includegraphics[width=8.5cm]{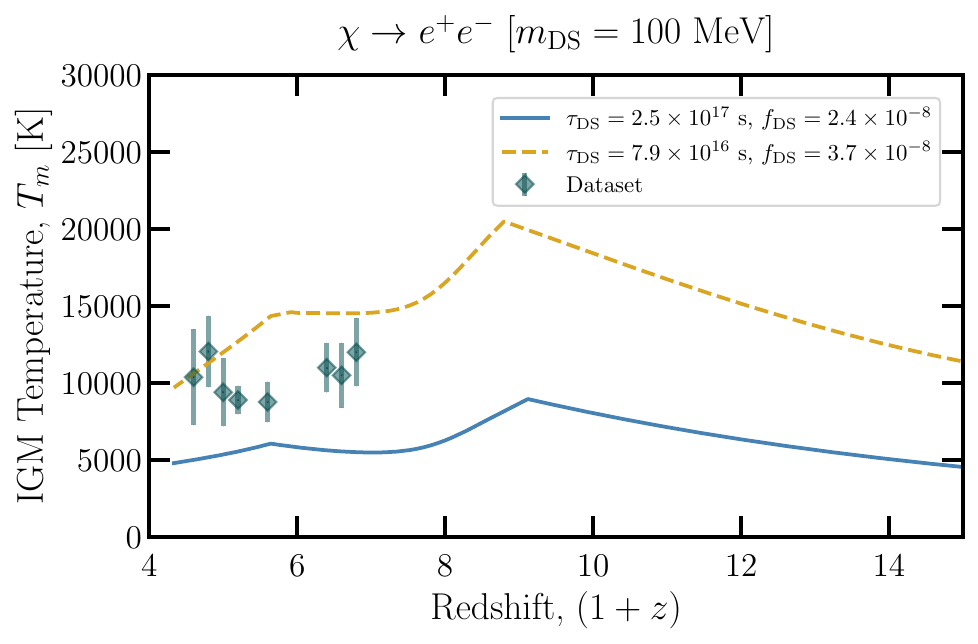}
    \caption{\textit{Left:} Ionization history showing the free electron fraction as a function of redshift for energy injections from the decays of DS of mass $m_{\rm DS} =$ 500 MeV, into $e^+e^-$. The solid red curve represents the tanh model for free electron fraction reproducing the $\textit{Planck 2018}$ constraints on $\tau_{\rm reion}$~\cite{Planck:2018vyg}. The DS parameters used for the three ionization histories are:  $\tau_{\rm DS} = 10^{13}$ s, $f_{\rm DS} = 4 \times 10^{-9},$ (solid blue); $\tau_{\rm DS} = 10^{15}$ s, $f_{\rm DS} = 4 \times 10^{-11}$ (yellow dot dashed); and $\tau_{\rm DS} = 10^{17}$ s, $f_{\rm DS} = 1.3 \times 10^{-10}$ (dotted), with the ionization level set to the tanh one at low redshifts. \textit{Right:} IGM temperature $T_{\rm m}$ as a function of redshift for two DS decay scenarios with $m_{\rm DS}=100\,\mathrm{MeV}$ and decay to $e^+e^-$. The dashed yellow curve corresponds to a case that overheats the IGM relative to Ly$\alpha$--inferred temperatures~\cite{Walther:2018pnn, Gaikwad:2020art} and is therefore excluded, while the solid blue curve remains under-heated and is consistent with the data. The Ly$\alpha$ measurements of $T_{\rm m}$ (shown in teal) constitute the fiducial dataset~\cite{Liu:2020wqz} used in this analysis. The ionization history is fixed to a tanh parametrization at low redshifts for $T_{\rm m}$ evolution with redshift.
}
    \label{fig:xe_compare}
\end{figure*}

For short-lifetime DS decays to $e^+e^-$ with $\tau_{\rm DS} =$  $10^{13}$ s, the deposition efficiency is strongly suppressed at low redshifts with respect to values at larger redshifts as can be seen in Fig.~\ref{fig:fc_two_column} (first column). For longer lifetime DS decaying to $e^+e^-$ with $\tau_{\rm DS} =$  $10^{17}$ s, we can moreover see that the deposition efficiency is highest for injection energies close to $10^8$ eV~\cite{Slatyer:2016qyl}. In all cases, we see that the energy efficiency functions are always  $f_c(z)< 1$ and depend strongly on the injection energies and redshift as well as deposition channel. In contrast to this, Ref.~\cite{Lucca:2019rxf} defined the deposition function as $f_c = f_{\rm eff} \ \chi_c(x_e)$ where $\chi_c(x_e)$ is the partition fraction into any channel (which they take from ~\cite{Galli:2013dna}), and  $f_{\rm eff}$ is set to 1. For the hydrogen ionization channel for example, $\chi_{\rm Hion}(x_e)$ ranges between $0.35-0.001$ which as we can see from Fig.~\ref{fig:fc_two_column} does not capture the full behaviour of these functions. This motivates revisiting the {\it Planck 2018} constraints for a self-consistent comparison with the  Ly$\alpha$ bounds, which we derive in this paper. 

For the case of DS decaying to $\gamma \gamma$ instead, the efficiency function for hydrogen ionization channel peak for DS mass close to $30-40$ eV~\cite{Slatyer:2016qyl}.

Using these efficiency functions defined in Eq.~\eqref{eq:fc_eqn}, the impact of the DS energy injection on the thermal and ionization history (see Eq.~\eqref{eq:evol_eq}) can be now written using the following terms
\begin{align}
\dot{T}_m^{\rm DS} &=
\frac{2\,f_{\rm heat}(z,\mathbf{x})}
{3 n_{\rm tot}}
\left(\frac{dE}{dV\,dt}\right)^{\rm DS, \  inj},
\\[6pt]
\dot{x}_{\rm HII}^{\rm DS} &=
\left[
\frac{f_{\rm H\,ion}(z,\mathbf{x})}{E_{\rm ion} \,n_{\rm H}}
+ \frac{(1- \mathcal{C})\,f_{\rm exc}(z,\mathbf{x})}{E_{\rm ly\alpha}\,n_{\rm H}}
\right]
\left(\frac{dE}{dV\,dt}\right)^{\rm DS,  inj}. \label{eq:xe_ds}
\end{align}

Here, $n_{\rm tot}$ is the total baryonic number density given by $n_{\rm H}(1 + x_e  + n_{\rm He}/n_{\rm H})$. In Eq.~\eqref{eq:xe_ds}, $\mathcal{C}$ is the Peebles-C factor or the probability of a hydrogen atom in the $n = 2$ state decaying to the ground state before photoionization can occur, $E_{\rm ly\alpha} = 10.2$ eV is the energy for Lyman-$\alpha$ excitation of hydrogen and $E_{\rm ion} = 13.6$ eV. The DS ionization term for the case of HeII can be written in an analogous way (see also Ref.~\cite{Liu:2019bbm}).~\footnote{Note that the DS contribution to doubly ionized Helium is set to 0, i.e., $\dot{x}_{\rm HeIII}^{\rm DS} = 0$~\cite{Liu:2019bbm} in \texttt{DarkHistory} and we follow this prescription in our analysis. Moreover, Ref.~\cite{Liu:2020wqz} studied the impact of neglecting this for the case of decaying DM constraints and found it to be less than 1\%, so we do not expect a substantial change for the DS case either.} \\[0.1cm]

The increased ionization level from DS energy injection can lead to a positive feedback on the efficiency functions $f_c(z, \mathbf{x})$ by the backreaction effect. The publicly available code \texttt{DarkHistory} allows the user to easily account for this while computing IGM evolution (see discussion in \cite{Liu:2019bbm}). Including this effect leads to a higher prediction of IGM temperature at low redshift $z \sim 5-10$, giving stronger bounds as we will discuss in the next section~\ref{sec:method}.

In Fig.~\ref{fig:xe_compare}, we show the free electron fraction $x_e$ and matter temperature $T_{\rm m}$ as a function of redshift for different scenarios of energy injection from a DS particle decaying to $e^+e^-$. In the left plot of Fig.~\ref{fig:xe_compare}, in addition to the curve (solid red) showing the tanh parametrization for free electron fraction $x_{e}(z)$ from Planck's constraints, we show three different cases of energy injection from DS decays,  assuming a decaying DS particle with mass $m_{\rm DS}=500\,\mathrm{MeV}$  decaying to $e^+e^-$, and the subsequent impact on $x_e$ between recombination at $z \sim 1100$ and the onset of reionization at $z \sim 10$. For $\tau_{\rm DS} = 10^{13}$ s (in blue solid) the largest change in $x_e$ is seen close to $z \sim 700$, close to recombination. The energy deposition efficiency for hydrogen ionization channel for a DS particle with such short lifetimes and decaying to $e^+e^-$ dominates at large redshifts as can be seen from Fig.~\ref{fig:fc_two_column} (top, leftmost). For $\tau_{\rm DS} = 10^{15}$ s (dot-dashed yellow), change in $x_e$ is dominant for $z \sim 60$, while for late decaying DS particle with lifetime $\tau_{\rm DS} = 10^{17}$ s (dotted teal), the large change in $x_e$ is visible at much late redshifts $z \sim 10$ due to later deposition of energy into the hydrogen ionization channel. In the right plot of Fig.~\ref{fig:xe_compare}, we instead show the evolution of IGM temperature for the case of DS decaying to $e^+e^-$  with  $m_{\rm DS}=100\,\mathrm{MeV}$, and the subsequent impact on $T_{\rm m}$, along with  IGM temperature measurements indirectly measured from Ly$\alpha$ forest observations~\cite{Walther:2018pnn, Gaikwad:2020art}. Here, $\tau_{\rm DS} =2.5\times 10^{17}$ s is shown with a dot-dashed yellow line, while $\tau_{\rm DS} =7.9\times 10^{16}$ s is shown with a blue solid line.

\noindent \textbf{Reionization models.} At redshifts below $z < z^*$, where $z^*$ denotes the redshift for the onset of reionization, injection from astrophysical sources driving reionization introduces many uncertainties. For details on this, see e.g. discussions in Refs.~\cite{Capozzi:2023xie, Liu:2020wqz, Poulin:2015pna, Liu:2019bbm} . \\
We adopt a conservative approach for accounting for IGM evolution at such low redshifts and use the Planck constraints on the optical depth to reionization $\tau_{\rm reion}$. The optical depth is computed as
\begin{align}\label{eq:optical_depth}
    \tau_{\rm reion} = \int_0^{z_{e,  {\rm  min}}}  dz \ n_e(z) \ \sigma_T \ \frac{dt}{dz} \ ,
\end{align}
where  $n_e(z) =x_e(z) \cdot n_{\rm H}$ is the redshift-dependent electron number density, and $\sigma_T$ is the Thompson cross-section for scattering. The upper limit of the integral has been set to $z_{e, {\rm min}}$,  the redshift at which the $x_e$ reaches a minimum value.\footnote{For the scenarios considered in our analysis $z_{e, {\rm min}}<750$.} This was shown to provide a robust definition of the optical depth for scenarios leading to an early enhancement of the ionized fraction in~Ref.~\cite{Poulin:2015pna}, see their appendix D.\\
The tanh model adopted in Planck analyses~\cite{Planck:2015fie} is based on a smooth transition from a neutral
to ionized Universe, where the parametric form for $x_e(z)$ is based
on a hyperbolic tangent (see also the solid red curve in Fig.~\ref{fig:xe_compare} showing $x_e(z)$ from the tanh model we use)~\footnote{The tanh model is defined using the following redshift dependence~\cite{Planck:2018vyg} 
\begin{equation}
x_e(z) =
\frac{1 + n_{\mathrm{He}}/n_{\mathrm{H}}}{2}
\left[
1 + \tanh\!\left(
\frac{y(z_{\mathrm{re}}) - y(z)}{\Delta y}
\right)
\right] ,
\end{equation} where $y(z) = (1 + z)^{3/2}$, $\Delta y = \frac{3}{2} (1 + z_{\mathrm{re}})^{1/2} \cdot 0.5$, and $z_{\mathrm{re}}$ is the measured reionization redshift.}. \\
Assuming the tanh model, the  optical depth measurements along with the 68\% CL error from the \textit{Planck 2018} dataset~\cite{Planck:2018vyg} is
\begin{align}\label{eq:tau_2015}
    \tau_{\rm reion}^{\rm Pl} = 0.054\pm 0.007 ,
\end{align}
while the \textit{Planck 2015}~\cite{Planck:2015fie} measurements instead give
\begin{align}\label{eq:tau_2018}
    \tau_{\rm reion}^{\rm Pl} = 0.066\pm 0.016 .  
\end{align}

\label{sec:method}
\begin{figure*}
\includegraphics[width=8.6cm]{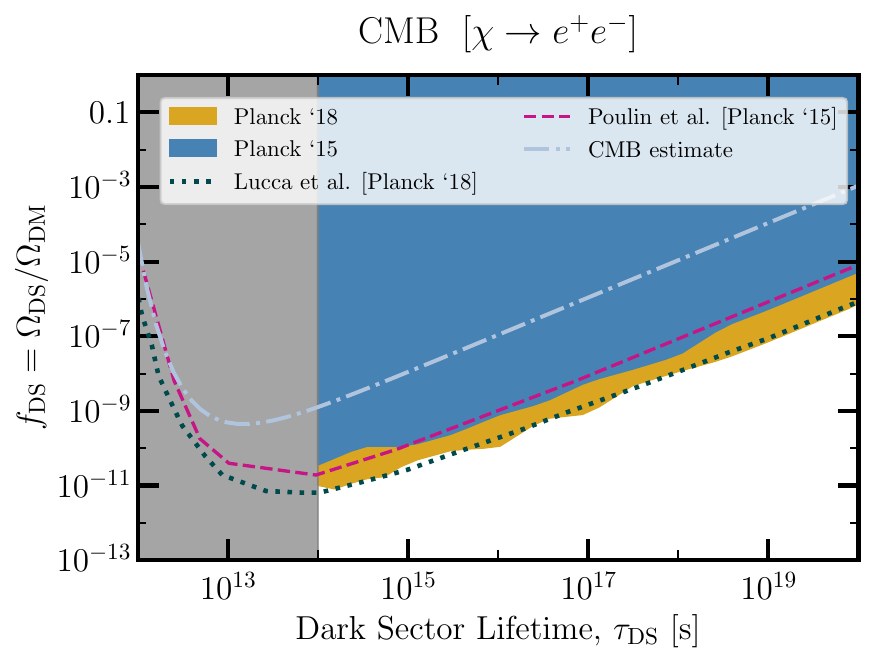}
\includegraphics[width=8.6cm]{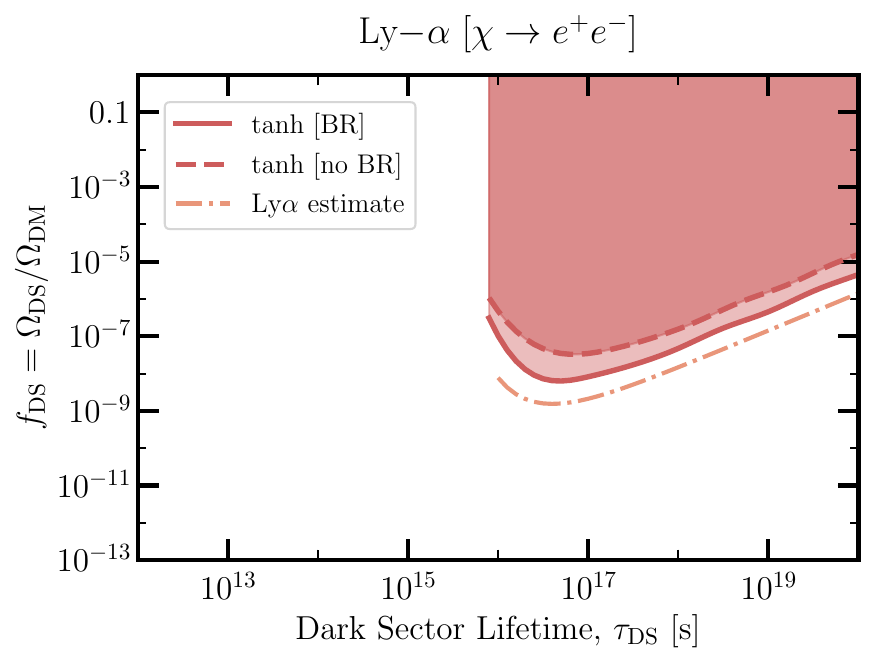}
    
    \caption{\emph{Left:} Constraints on DS parameter space obtained using CMB estimates based on the evaluation of the optical depth to reionization (see main text) for DS particle $\chi$ decaying to $e^+e^-$. The  blue and yellow regions are the most stringent limits we obtain in this work using the reported values of $\tau_{\rm reion}$ in \textit{Planck 2015}~\cite{Planck:2015fie} and \textit{Planck 2018}~\cite{Planck:2018vyg} datasets for DS mass $= 100$ MeV. We further compare our results with the strongest existing bounds on decaying dark sectors obtained in previous studies, derived from full CMB analyses of the Planck 2015 data by Poulin et al.~\cite{Poulin:2016anj} (dashed pink), and of the Planck 2018 data by Lucca et al.~\cite{Lucca:2019rxf} (dotted teal), where in the latter case, the authors have used a constant efficiency function $f_{\rm eff} = 1$ (see discussion in the main text). The vertical grey region indicates the region of $\tau_{\rm DS} \lesssim 10^{14}$ s where our CMB bounds based on optical depth reionization cannot constrain ionization level change (see discussion in main text). Our estimated CMB bounds using Eq.~\eqref{eq:stima_xe} are also shown for comparison as a dot-dashed line.  \emph{Right:} Constraints on DS parameter space obtained from IGM heating using Ly$\alpha$ temperature measurements for DS with mass $m_{\rm DS} = 10^{8.5}$eV, decaying to $e^+e^-$, where we have set the ionization free electron fraction for low redshifts to Planck's tanh model. Here the dashed red curve shows the bounds without backreaction [no BR] and solid red shows the stronger bounds with backreaction [BR] included. In pink dot-dashed curve, the estimated bound for Lyman$-\alpha$ using Eq.~\eqref{eq:stima_Tm} are shown. } 
    \label{fig:cmb_ee}
\end{figure*}

\section{method and constraints}
\label{sec:method}
\noindent \textbf{CMB bounds.} As previously discussed, energy injections from generic dark sector decays can leave an observable imprint on CMB anisotropies, leading to stringent constraints on the fractional abundance $f_{\rm DS}$ of such metastable decaying DS particles. \\
In this section, we give details on obtaining our CMB bound estimated using the reionization optical depth  measurements from \textit{Planck 2015} and \textit{Planck 2018} datasets (see Eqs.~\eqref{eq:tau_2015} and \eqref{eq:tau_2018}). Note that in regions of the parameter space where DS decays are mostly affecting reionization (and not recombination), this method is expected to provide equivalent 95\% CL bounds as a full MCMC analysis taking into account the full  \textit{Planck 2018} dataset and the degeneracies with other cosmological parameters, see e.g. Ref.~\cite{Capozzi:2023xie} for a concrete check within an ALP DM scenario.  \\
The estimated bounds are computed in this work in the following way. Using the modified version of \texttt{DarkHistory} code as well as the obtained efficiency functions for DS decays, we evaluate the optical depth to reionization using Eq.~\eqref{eq:optical_depth} over a range of DS parameter $(\tau_{\rm DS}, f_{\rm DS})$. We then estimated the CMB bounds by excluding the region where $\tau_{\rm reion}(\rm model) > \tau_{\rm reion}^{\rm Pl} + 2  \cdot \sigma_{\rm Pl}$ with the $\tau_{\rm reion}^{\rm Pl}$ and $\sigma_{Pl}$ for the Planck 2018 and 2015 datasets given respectively by Eq.~\eqref{eq:tau_2015} and Eq.~\eqref{eq:tau_2018}. 

We show the most stringent exclusions obtained on the DS parameter space for the case of DS decaying to $e^+e^-$ for DS mass $m_{\rm DS} = 100$ MeV, in Fig.~\ref{fig:cmb_ee} as blue and yellow exclusion regions respectively for the Planck 2015 and 2018 cases. We further compare our estimated bounds with the most stringent full MCMC CMB analyses limits derived using the Planck 2015 and 2018 datasets in Refs.~\cite{Poulin:2016anj, Lucca:2019rxf}. Our estimates capture well the behavior of Planck 2015 and 2018 constraints derived using the full CMB analysis for the $\tau_{\rm DS} \gtrsim 10^{14}$ s. 
Compared to ref.~\cite{Lucca:2019rxf}, which assumed a simplified DS decaying model with $f_{\rm eff}=1$, our constraints in the long-lifetime regime (with $\tau_{\rm DS} > 10^{14}$ s) are nearly comparable.
Note however that the limits in ~\cite{Lucca:2019rxf} are for decays to all electromagnetic particles including $\gamma$s and $e^+e^-$ injection whereas in deriving our bounds in Fig.~\ref{fig:cmb_ee}, we are considering only $e^+e^-$ primary injected particles with DS mass set to 100 MeV.  \\
The DS bounds reported in Ref.~\cite{Poulin:2016anj} for $e^+e^-$ injection from DS decay, shown as a dashed magenta curve in Fig.~\ref{fig:cmb_ee}, correspond to their strongest constraints without the backreaction effect. The small differences between their results and ours can be attributed to updates in the energy deposition efficiency functions $f_c(z)$, as well as to our inclusion of the backreaction effect.\\
For shorter lifetimes, $\tau_{\rm DS} \lesssim 10^{14} \ \mathrm{s}$, energy injection from DS decays occurs close to recombination (for illustration see the blue curve in the $x_e$ plot in Fig.~\ref{fig:xe_compare}). In this regime, the resulting change in $x_e$ is not well captured by an effective shift in the optical depth to reionization, and consequently our CMB bounds cannot reproduce the results of a full CMB anisotropy analysis, as performed in Refs.~\cite{Lucca:2019rxf, Slatyer:2016qyl, Poulin:2016anj, Acharya:2019uba}. Our bounds based on the evaluation of $\tau_{\rm reion}$ are only sensitive to the integrated effect of DS energy injection closer to reionization.  While the main goal of our work is to compute the impact of DS heating on the IGM and constrain it using Ly$\alpha$ IGM measurements, the main motive behind this CMB estimate is to have a self-consistent comparison with the current CMB limits (Planck 2018), taking into account the energy deposition functions $f_c(z)$ and further including backreaction effect for the same lifetime range ($\tau_{\rm DS}>10^{14}$ s).

For the decay of DS to $\gamma \gamma$ final states, our bounds are illustrated in Fig.~\ref{fig:cmb_mds} (right plot) in appendix~\ref{app:cmb_mds}, for DS mass values going from 40 eV to 1 TeV, with the strongest limits derived for $m_{\rm DS} = 40$ eV.  In the left plot of Fig.~\ref{fig:cmb_mds}, we also show how the CMB bounds for DS decaying to $e^+e^-$ states change by varying the DS mass. For this case, the strongest bounds from CMB are obtained in the DS mass range $\sim 30 - 300$ MeV, due to the large deposition efficiency for hydrogen ionization, $f_{\rm H ion}$ (see top row in  Fig.~\ref{fig:fc_two_column} where $f_{\rm H ion} > 0.1$ for $K_{\rm inj} \sim 10^8$ eV). This increase in $f_{\rm H ion}$ for energies close to 100 MeV is due to  the fact that electrons in this energy range can efficiently upscatter CMB photons to $\sim 10 - 10^3$ eV energies, which can subsequently ionize hydrogen in the IGM. For injection of electrons with lower and higher energies, the resulting upscattered CMB photons are not efficient ionizers~\cite{Slatyer:2016qyl, Lopez-Honorez:2016sur}. Instead, for the case of $\chi \to \gamma \gamma$ decay, the energy deposition efficiency function for hydrogen ionization peaks for much lower values of injected energies close to 30 - 40 eV~\cite{Slatyer:2016qyl, Agius:2025nfz}. \\
We further note that in comparison to \cite{Poulin:2016anj} which consider injection energies down to only 10 keV for DS decaying to photons, we consider much lower injection energies by deriving bounds for DS mass down to 40 eV for the $\chi \to \gamma \gamma$ case. \\[0.2cm] 

\noindent \textbf{Ly$\alpha$ bounds.} Late-time energy injections from DS decays can lead to excessive heating of the IGM. Such DS decays can be constrained with existing and forthcoming temperature measurements from Ly$\alpha$ forest spectra and upcoming 21-cm power spectrum measurements. In this work, we focus on the former probe. Using our modified version of \texttt{DarkHistory} code and the computed efficiency functions for DS decays, we obtain IGM temperature histories over a range of DS model parameters $(\tau_{\rm DS}, f_{\rm DS})$. For our conservative bounds, we neglect the astrophysical sources of heating at $z < z^*$ (setting $\dot{Y}^{\rm astro} = 0$ in Eq.~\eqref{eq:evol_eq}) and fix the ionized fraction at these redshifts to Planck's tanh model~\cite{Planck:2018vyg}. 
We then compare the obtained IGM model predictions to dataset of IGM temperatures based on Ly$\alpha$ data in Refs.~\cite{Liu:2020wqz,  Walther:2018pnn, Gaikwad:2020art}. \\
Refs.~\cite{Walther:2018pnn, Gaikwad:2020art} obtained their IGM temperature measurements by comparing hydrodynamical simulation generated mock Ly$\alpha$ spectra and comparing it to observed spectra of quasars obtained from spectroscopic surveys. This dataset we use following Ref.~\cite{Liu:2020wqz} (see also Fig.~\ref{fig:xe_compare} right plot) is based on a combination of IGM temperature measurements obtained from Walther et al.~\cite{Walther:2018pnn} in the range $4.6 <(1 + z)< 5.6$ and Gaikwad et al.~\cite{Gaikwad:2020art} within $6.4 < (1+z) < 6.8$, where we have taken only the data above $z > 3.6$ such that the redshifts are away from the redshifts associated to full HeII reionization to HeIII, allowing us to use \texttt{DarkHistory} which assumes $\dot{x}_{\rm HeIII} = 0$. \\

For excluding DS parameters $(\tau_{\rm DS}, f_{\rm DS})$ causing excessive IGM heating, we perform the modified $\chi^2$ test based on Ref.~\cite{Liu:2020wqz}. In Fig.~\ref{fig:cmb_ee} (right), we show the results from our analysis for finding the DS parameters that are
excluded by the Ly$\alpha$ data at 95\% CL where the DS mass is fixed to the value giving the most stringent constraints ($m_{\rm DS} = 10^{8.5}$ eV) for DS decay to $e^+e^-$. The dashed red and solid red curves in Fig.~\ref{fig:cmb_ee} (right) show the bounds we obtain for the fixed tanh reionization model where the effects of backreaction have been neglected and taken into account respectively. As already shown in Ref.~\cite{Liu:2019bbm} for any energy injection, the most important impact of including the backreaction effect is in the increase in the energy deposition into IGM heating which leads to a much higher prediction of the IGM temperature at late redshifts, and thus more stringent bounds when including the effect (as can be seen in Fig.~\ref{fig:cmb_ee} right plot). 

In Fig.~\ref{fig:lyman_mds} (Appendix~\ref{app:lyman_mds}), we instead show, for both DS decays into \(e^+e^-\) and \(\gamma\gamma\) final states, the impact of varying the DS mass on the resulting exclusion limits. We present the cases corresponding to the DS mass values that yield the strongest and weakest IGM heating effects over the redshift range probed by Ly\(\alpha\) forest measurements. For the $\chi \to e^+e^-$ case (Fig.~\ref{fig:lyman_mds}, left), the best constraints are obtained for $m_{\rm DS} \sim 300$ MeV with the weakest limits obtained for $m_{\rm DS} \sim 1$ TeV. This is attributed to the large value of deposition efficiency for IGM heating channel obtained for DS mass $\sim 100$ MeV at the relevant redshift for Ly$\alpha$ measurements, $f_{\rm heat}(z=5) \sim 0.3$. The deposition function value for DS mass $\sim 1$ TeV, is instead much smaller $\sim \mathcal{O}(10^{-3})$ for the same redshifts.
For the $\chi \to \gamma\gamma$ decay channel (Fig.~\ref{fig:lyman_mds}, right), the strongest bounds are obtained for $m_{\rm DS} = 100 \ \mathrm{eV}$. We additionally show bounds for DS masses of $100 \ \mathrm{MeV}$, 40 eV, and 10 keV, out of which the weakest bounds are given by $m_{\rm DS} = 100$ MeV. The bounds for DS mass of $1 \ \mathrm{TeV}$ in the $\chi \to \gamma\gamma$ channel are not shown in Fig.~\ref{fig:lyman_mds}, as they are much weaker, by an additional factor of $\sim 30$ with respect to the $100 \ \mathrm{MeV}$ case.

\begin{figure*}
\includegraphics[width=8.5cm]{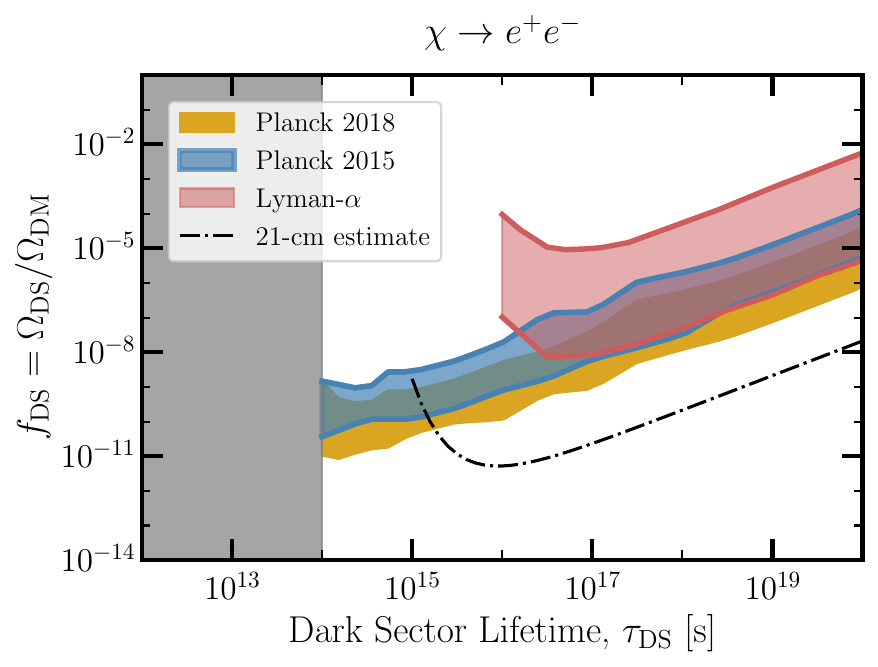}
\hfill
\includegraphics[width=8.5cm]{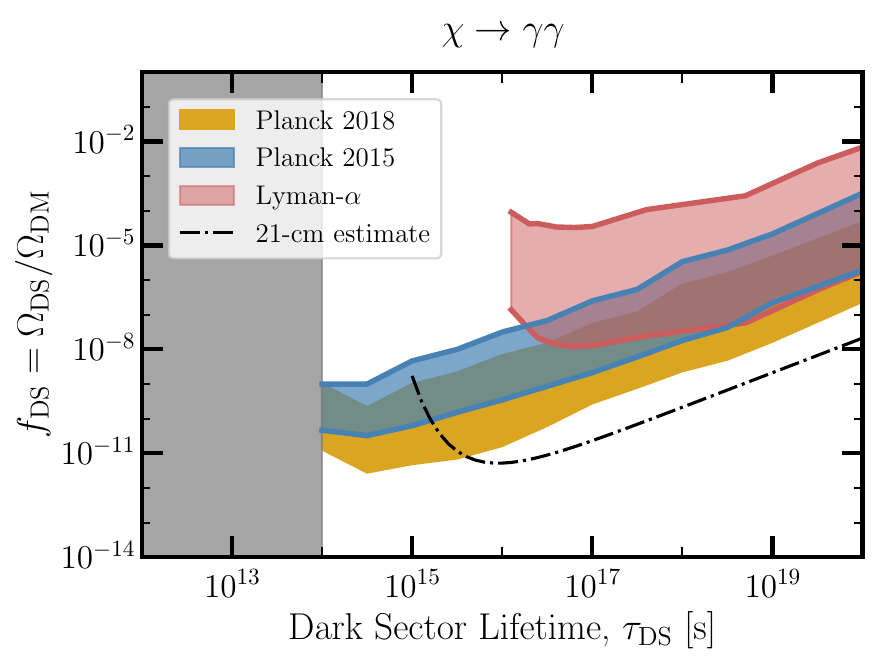}
\caption{\emph{Left:} Constraints derived in this work for DS decaying to $e^+e^-$ for the ionization history fixed to tanh model. The width of the bands correspond to the extreme assumptions on the DS mass from 1 TeV (weakest) to 316 MeV (strongest) for Ly$\alpha$ bounds and in the range 1 TeV - 100 MeV for the CMB bounds. The red band has been derived using Lyman-$\alpha$ IGM temperature measurements, whereas the blue and yellow bands have been derived using CMB optical depth evaluation for the \emph{Planck 2015} and \emph{Planck 2018} measurements respectively. The vertical grey region indicates the region of $\tau_{\rm DS} \lesssim 10^{14}$ s where our CMB bounds based on optical depth reionization cannot constrain ionization level change (see discussion in main text). We additionally show the estimated bounds using Eq.~\eqref{eq:stima_Tm} that will be probed by 21-cm probes (dot-dashed). \emph{Right:} Same constraints derived in this work for DS decaying to $\gamma \gamma$ where for the Lyman$-\alpha$ band (red), we vary the DS mass between 100 MeV (weakest) to 100 eV (strongest) and for the CMB constraint bands, between 1 TeV (weakest) and 39.8 eV (strongest).}
\label{fig:all_cons}
\end{figure*}
\vspace{0.4cm}
\noindent \textbf{Summary plots.} In Fig.~\ref{fig:all_cons}, we summarize all the constraints derived in this work using CMB optical depth evaluation and Lyman$-\alpha$ IGM temperature for DS decaying to $e^+e^-$ and $\gamma \gamma$. Our bounds are shown as red (Lyman-$\alpha$), blue (Planck 2015), and yellow (Planck 2018) bands where the strongest and weakest limits are obtained by varying the DS mass in each case (see Figs.~\ref{fig:cmb_mds} and ~\ref{fig:lyman_mds}). From Fig.~\ref{fig:all_cons} for DS decaying to $e^+e^-$, we see that the bounds derived using Ly$\alpha$ are competitive with the CMB Planck 2015 bounds for longer DS lifetime close to $ \tau_{\rm DS} \gtrsim 10^{17} \ \mathrm{s}$.  However, the limits derived using latest CMB Planck 2018 optical depth measurement (yellow band in Fig.~\ref{fig:all_cons}) are the strongest obtained. For the DS decaying to $\gamma \gamma$ case (right plot in Fig.~\ref{fig:all_cons}) we find that the Ly$\alpha$ limits are stronger than the bounds obtained from Planck 2015 optical depth measurements for $\tau_{\rm DS} \gtrsim 5 \times 10^{18}$ s, however, those from Planck 2018 are eventually the strongest of all for the entire DS lifetime range. Finally, using the estimate in  Eq.~\ref{eq:stima_Tm}, we also show future projection of 21-cm probes in Fig.~\ref{fig:all_cons}.\\   
Note that, in our analysis, the strongest CMB constraints for DS decays into $\gamma\gamma$ are more stringent than those for DS decays into $e^+e^-$. Unlike earlier studies (e.g. Ref.~\cite{Poulin:2016anj}), our photon energy injection probes much lower DS masses, down to $\sim 40 \ \mathrm{eV}$, whereas Ref.~\cite{Poulin:2015pna} considered injection energies ranging only from $1 \ \mathrm{TeV}$ down to $10 \ \mathrm{keV}$. See also Refs.~\cite{Capozzi:2023xie, Liu:2023nct} for discussion on injection of lower energies for photon final states.\\

\section{Conclusion and outlook}
\label{sec:summary}
In this work, we have derived constraints on decaying dark sectors using Ly$\alpha$ IGM temperature measurements and self-consistently revisited CMB bounds using Planck's measurements of the reionization optical depth. In order to compute our bounds, we have made use of the publicly available code \texttt{DarkHistory}~\cite{Liu:2019bbm, Sun:2022djj} by further modifying it to handle energy injections from a metastable DS particle with lifetime shorter than the age of the Universe. Using this extended code, we have  computed energy deposition functions for the channels for short-lifetime DS decaying to $e^+e^-$ and $\gamma \gamma$, reproducing results in literature~\cite{Slatyer:2016qyl}. In  our work we have  consistently  incorporated these energy  deposition functions $f_c(z, \mathbf{x})$  throughout our Ly$\alpha$ and CMB analyses.\\

In order to derive our cosmological bounds on the DS parameter space from IGM heating using Ly$\alpha$ data, we adopted a conservative approach where we have fixed the ionization history at low redshifts before the onset of reionization to the tanh model used in Planck's analysis, see, e.g.~\cite{Planck:2018vyg}. In this way, in deriving our final bounds from Ly$\alpha$ data, we have not considered any astrophysical source of photoheating. 

Our constraints from Lyman$-\alpha$ data are shown in Fig.~\ref{fig:all_cons} with a red band together with the CMB limits (blue and yellow bands) that we have revisited in this work. The width of the bands corresponds to our assumption on the DS mass, which, for the case of DS decay to \(e^+e^-\) (left plot in Fig.~\ref{fig:all_cons}), was scanned between 1~TeV and 100~MeV. For the CMB and Lyman-\(\alpha\) cases, the most stringent constraints are obtained for 100~MeV and 316~MeV, respectively, while the weakest limits are obtained for 1~TeV in both cases. For the DS decay to $\gamma \gamma$ case instead (right plot in Fig.~\ref{fig:all_cons}), to obtain the bands, we scanned DS masses between 100 MeV - 40 eV (1 TeV - 40 eV) for obtaining the Lyman$-\alpha$ bounds (CMB bounds). For the CMB and Lyman-$\alpha$ cases, the most stringent limits are obtained for DS masses of 40 eV and 100 eV, respectively, whereas the weakest limits obtained correspond to 1 TeV and 100 MeV. 
This further shows that different cosmological probes are sensitive to different injection energies -  demonstrating the complementarity of these probes.\\ For the DS decay to $e^+ e^-$ (right plot in Fig.~\ref{fig:all_cons}), we find that our derived Ly$\alpha$ limits can exclude DS fractional abundance close to $f_{\rm DS} \sim 8 \times 10^{-9}$ in the long-lifetime regime $\tau_{\rm DS} \sim 5 \times 10^{16}$ s. These are competitive (weaker) than the CMB Planck 2015 (2018) bounds from the optical depth to reionization that we have derived in this work for the same range of lifetimes.\\ The main difference in our Planck 2015 and 2018 CMB bounds in Fig.~\ref{fig:all_cons} is attributed to the difference in optical depth measurements $\tau_{\rm reion}^{\rm Pl}$ and the variance $\sigma_{\rm Pl}$ between the two datasets (see Eq.~\eqref{eq:tau_2015} and Eq.~\eqref{eq:tau_2018}). In particular,  the constraints from Planck 2018 optical depth measurement exclude $f_{\rm DS} \gtrsim 7 \times 10^{-10}$ for  $\tau_{\rm DS} \sim 5 \times 10^{16}$s (comparable CMB bounds were obtained in  Ref.~\cite{Lucca:2019rxf} assuming an effective energy deposition $f_{\rm eff}=1$ and ignoring backreaction).\\ 
For the DS decay to $\gamma \gamma$ (left plot in Fig.~\ref{fig:all_cons}), our derived limits from Lyman$-\alpha$ data are slightly stronger with respect to those derived from Planck 2015 for $\tau_{\rm DS} \gtrsim 10^{18}$ s.  However, the strongest limits are those obtained from the Planck 2018 optical depth measurement. The latter exclude DS fractional abundance close to $f_{\rm DS} \gtrsim 10^{-11}$ for $\tau_{\rm DS} \sim 2 \times 10^{14}$ s.

Note, that the CMB constraints derived here fail to capture $x_e$ changes near recombination and cannot be used to constrain  DS lifetime $\tau_{\rm DS} \lesssim 10^{14}$  s (see gray area in  Fig.~\ref{fig:cmb_ee} and Fig.~\ref{fig:all_cons} indicating the limit of validity of our CMB bounds). A full CMB analysis would be needed to better capture the change in ionization level due to energy injection at $z \sim 1100$ (corresponding to $\tau_{\rm DS} \sim 10^{13}$ s). This is not the purpose of this work and the revisited CMB bounds are only provided to allow for a fair comparison to the conservative Ly$\alpha$ bounds derived here in a similar range of lifetimes ($ \tau_{\rm DS} > 10^{14}$ s). We further note that the results obtained here indicate that more advanced Ly$\alpha$ analyses, accounting for both astrophysical sources of heating and exotic energy injection, could easily improve upon current cosmological bounds from the CMB. Furthermore, 21 cm power spectrum measurements, which are sensitive to the IGM temperature at $z \sim 5- 20$, can be expected to significantly improve upon current cosmological probes by up to two orders of magnitude for decays into $e^+e^-$ (see the dot-dashed lines in Fig.~\ref{fig:all_cons}). 

As a further outlook to this work let us briefly discuss 
an application of our derived limits to evaporating Primordial Black Holes.  In Fig.~\ref{fig:pbh_cons} in App.~\ref{app:pbh_cons}, we translate the DS bounds from CMB and Lyman$-\alpha$ we obtained in Fig.~\ref{fig:cmb_ee} to the case of evaporating PBHs and compare our translated bounds to PBH limits derived in literature. Our CMB bounds agree quite well with previous results (e.g. see~\cite{Poulin:2016anj}) while our Ly$\alpha$ bounds provided are again competitive bounds for PBH masses larger than $\sim 3 \times 10^{14}$ g. Our estimate of 21cm cosmology constraints are comparable within a factor $\lesssim 10$ with respect to the recent work of Ref.~\cite{Sun:2025ksr} probing $f_{\rm PBH} \gtrsim 10^{-12}$ for PBH mass close to $2 \times 10^{14}$ g. 
 
To summarise, in this work we have extended previous bounds on DS parameter space and demonstrated the power of late-time probes such as existing IGM temperature measurements from Lyman$-\alpha$ spectra of quasars, in probing energy injection at $z \ll 1000$. CMB anisotropies, as a probe of exotic energy injection, is a competitive probe of e.g. s-wave annihilation of DM~\cite{Planck:2018vyg}. For DM or DS decays affecting the ionization fraction $x_e$ at later time,  late-time cosmological probes (including Ly$\alpha$ data and 21cm cosmology) can be expected to be much more sensitive to injections at $z \sim 5 - 20$, see e.g.~\cite{Diamanti:2013bia, Facchinetti:2023slb, Sun:2022djj}. These bounds are further complementary to those from CMB spectral distortion and BBN which dominate at much earlier timescales constraining DS lifetimes as small as $\tau_{\rm DS} \sim 10^8$ s (see Ref.~\cite{Poulin:2016anj} for a review of these).  While our work already shows the potential of Lyman$-\alpha$ IGM measurements to extend current CMB limits, upcoming 21-cm power spectrum measurements such as the ones that will be made by HERA~\cite{HERA:2021bsv} and SKA~\cite{inproceedings} can be expected to improve the current limits in the long DS lifetime regime by upto two orders of magnitude (see also our Fig.~\ref{fig:all_cons} for these estimated projections with dot-dashed lines). We leave a full 21-cm analysis for deriving projections on the DS parameter space for a future work.
\appendix
\begin{figure*}
    \includegraphics[width=8.5cm]{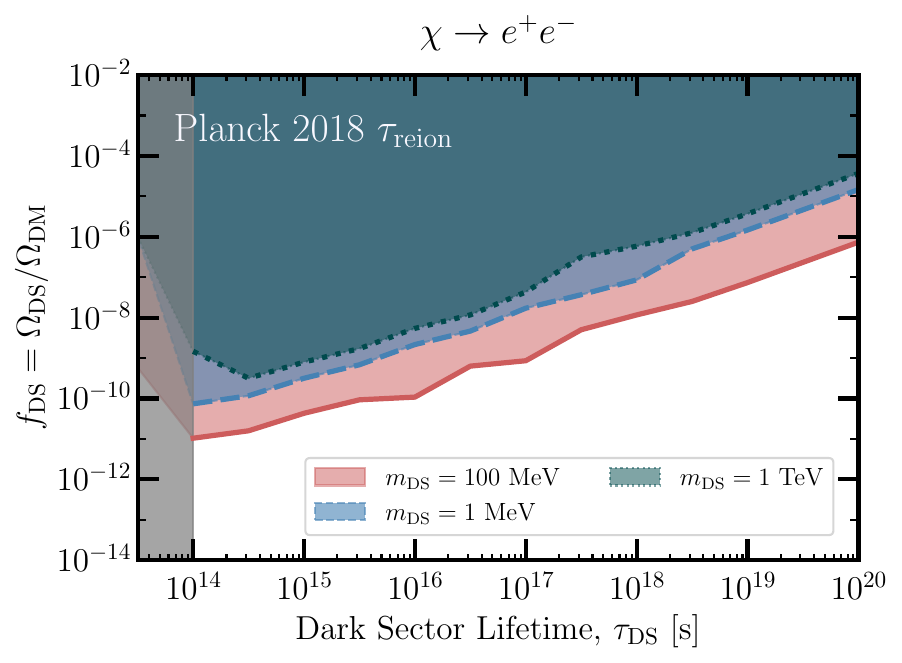}
    \hfill
    \includegraphics[width=8.5cm]{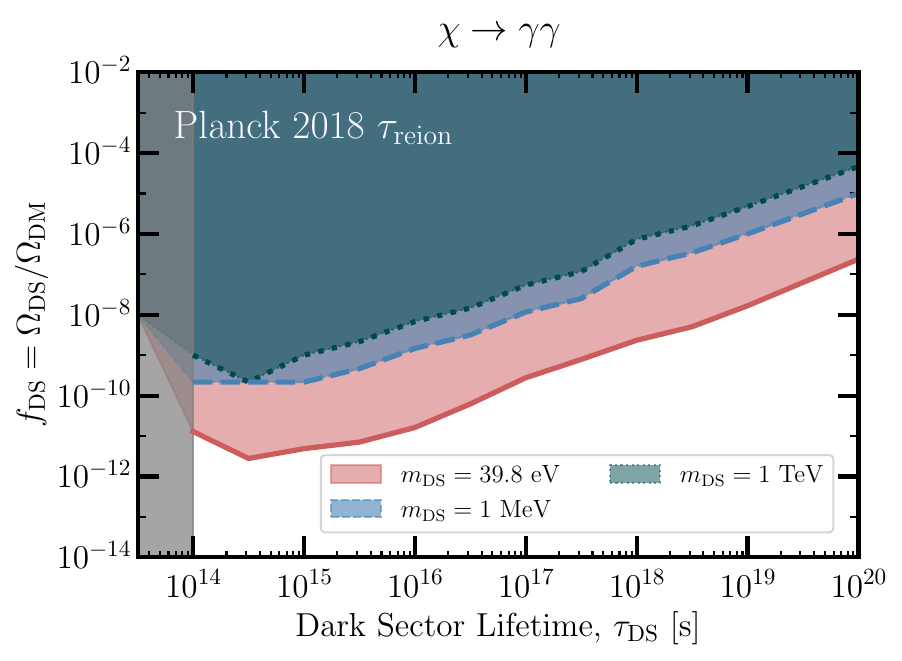}
    \caption{Constraints on DS parameter space obtained using CMB estimates based on the evaluation of the optical depth to reionization from Planck 2018 measurement of $\tau_{\rm reion}$ for DS decaying to $e^+e^-$ \textit{(left)} and for DS decaying to $\gamma \gamma$ \textit{(right)} final states for different DS mass values. For the $e^+e^-$ case, DS mass values in the range 30 - 300 MeV give some of the most competitive bounds, here we have shown the strongest that we obtain (for $m_{\rm DS}$ set to 100 MeV) to not crowd the plot. For $\gamma \gamma$ case, the strongest bounds are obtained for DS mass in the sub-keV range - the best we obtain is for DS mass close to 40 eV. }\label{fig:cmb_mds}
\end{figure*}

\section{Variation of CMB optical depth bounds with DS mass}\label{app:cmb_mds}
In this appendix, we show how the CMB bounds we obtained using our evaluation on the optical depth to reionization for Planck 2015 and 2018 cases vary with the DS mass for the $e^+e^-$ and $2 \gamma$ DS decay case.  Our resulting bounds for a range of $m_{\rm DS}$ can be seen in Fig.~\ref{fig:cmb_mds}.  

\section{Variation of Ly$\alpha$ bounds with DS mass}\label{app:lyman_mds}

\begin{figure*}
    \includegraphics[width=8.5cm]{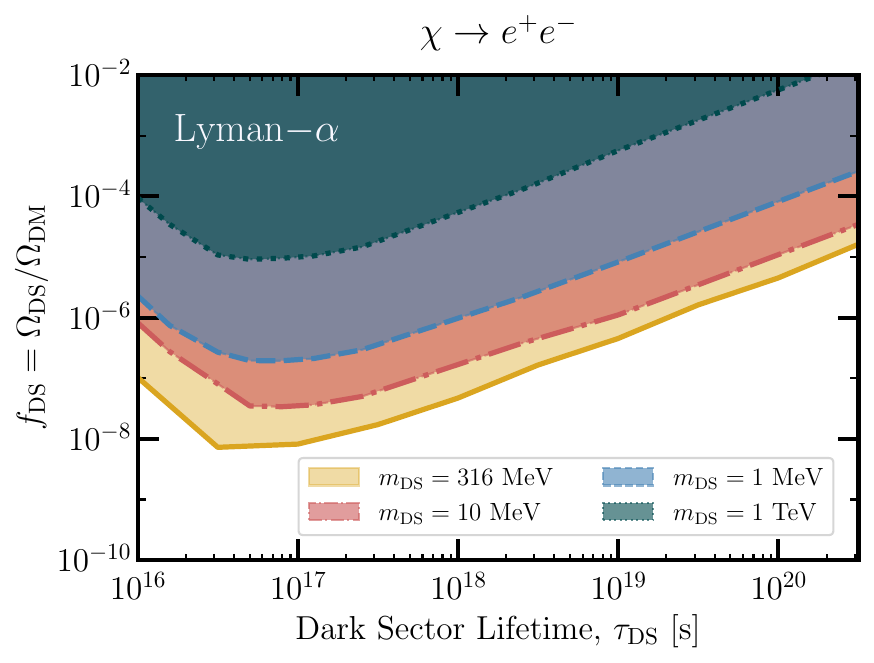}
    \hfill
    \includegraphics[width=8.5cm]{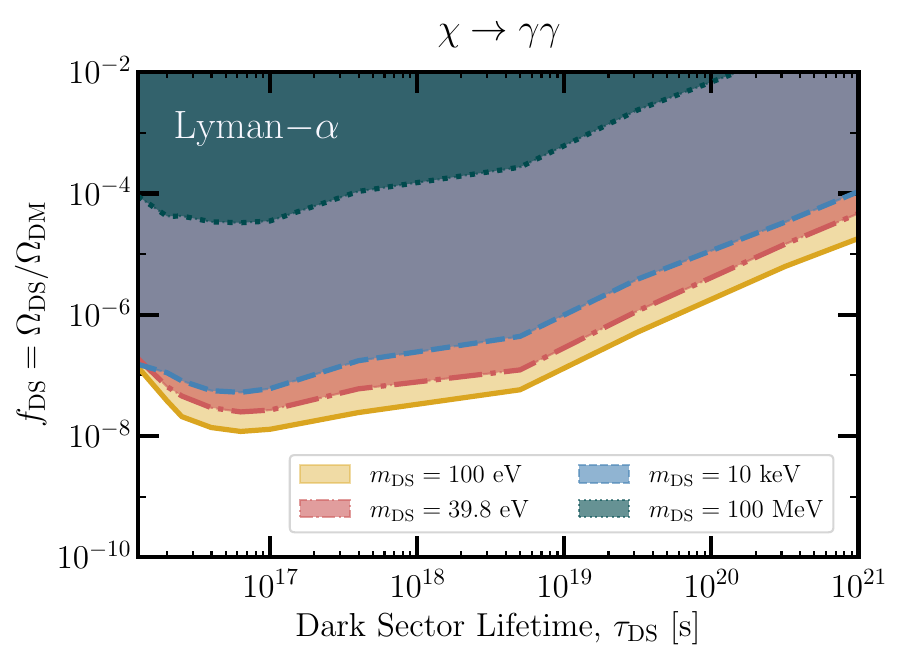}
    \caption{Constraints on DS parameter space obtained using Ly$\alpha$ IGM temperature measurements for a range of DS mass values going from 1 MeV to 1 TeV for DS decay to $e^+e^-$ case \emph{(left)} and from close to 40 eV upto 100 MeV for $\gamma \gamma$ \emph{(right)} case. The bounds have been derived using the fixed tanh ionization history and with the backreaction effect included for the two cases. The most competitive bounds are obtained for DS mass close to 300 MeV (left plot) for $e^+e^-$ case, and for DS mass of 100 eV (right plot) for the $\gamma \gamma$ case.}
    \label{fig:lyman_mds}
\end{figure*}
In this appendix, we show how the Lyman$-\alpha$ bounds we obtained using our analysis vary with the DS mass for the $e^+e^-$ DS decay case and the DS decaying to two photons case. 
The bounds can be seen in Fig.~\ref{fig:lyman_mds} where for the $e^+e^-$ case the most stringent constraints are obtained for $m_{\rm DS} \sim $ 300 MeV while for the $\gamma \gamma$ decay case, the best constraints are obtained for $m_{\rm DS} = 100$ eV as $f_{\rm heat}$ for the two decay channels peaks in these regions of injected energy~\cite{Slatyer:2016qyl, Liu:2019bbm, Agius:2025nfz}. 

\section{Application to PBH evaporation}\label{app:pbh_cons}

\begin{figure}[h!]
\includegraphics[width=0.5\textwidth]{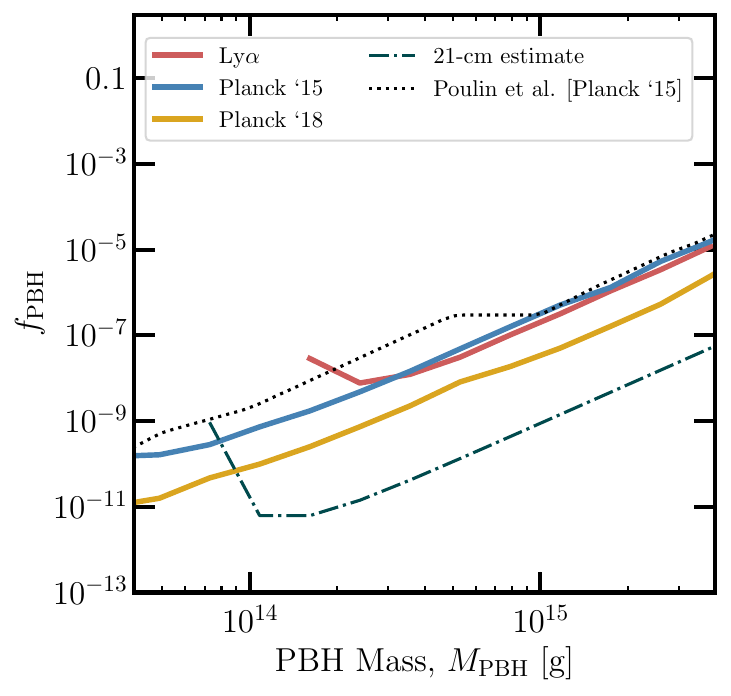}
  \caption{Translation of our derived bounds from CMB optical depth measurements from Planck 2015 (shown in blue) and Planck 2018 (shown in yellow) and Ly$\alpha$ data (shown in red) of Fig~\ref{fig:cmb_ee} to constraints on evaporating PBHs. In dotted black, we show PBH bounds derived in Ref.~\cite{Poulin:2016anj} from CMB analysis of Planck 2015 dataset.}
  \label{fig:pbh_cons}
\end{figure}

 Without performing further model-specific analyses (see Refs.~\cite{Poulin:2016anj, Lucca:2019rxf}), here we apply our derived bounds in Fig.~\ref{fig:cmb_ee} to a well-motivated decaying DS scenarios - the case of low-mass PBHs where energy injection occurs due to their evaporation.  For our translation, we use that the lifetime of an evaporating PBH is roughly given by~\cite{Carr:2009jm}
\begin{equation}
    \tau_{\rm PBH} \simeq 4.07 \times 10^{11} \bigg(\frac{\mathcal{F}(M)}{15.35} \bigg)^{-1} \bigg( \frac{M}{10^{13}\ \rm{g}} \bigg)^3 \ \rm{s},
\end{equation}

where $\mathcal{F}(M)$ counts the number of relativistic particles species emitted by the PBH. Our translated results are shown in the PBH parameter space spanned by the PBH mass $M_{\rm PBH}$ and PBH fraction $f_{\rm PBH}$ in Fig.~\ref{fig:pbh_cons}.\\ In comparison we show CMB bounds derived from energy injection with evaporating PBHs in Ref.~\cite{Poulin:2016anj} in dotted black. For $M_{\rm PBH} \gtrsim 10^{15}$g, our derived CMB bounds from optical depth estimate match the evaporating PBH constraints of Ref.~\cite{Poulin:2016anj} very well. For $M_{\rm PBH} \lesssim 10^{15}$g, our translated CMB bounds seem to overestimate the exclusions with respect to the bounds of Ref.~\cite{Poulin:2016anj}. This is due to the following reason: the kink below $10^{15}$ g is due to the effect of new channels like muon pairs opening up. Since these channels are less effective in releasing electromagnetic energy, they result in lowering the energy deposition efficiency, thereby, weakening the bounds. Since in obtaining Fig.~\ref{fig:pbh_cons}, we do not re-compute the efficiency functions for evaporating PBHs, our translated bounds do not capture this behaviour in the CMB bounds. \\
Recent analyses~\cite{Saha:2024ies, Khan:2025kag} have further derived bounds on evaporating PBHs from Ly$\alpha$ data, and their limits excluding roughly $f_{\rm PBH} \gtrsim 10^{-7}$ for $M_{\rm PBH} \sim 10^{15}$ g, are in the same ballpark as our Ly$\alpha$ translated bounds in Fig.~\ref{fig:pbh_cons}. We once again note that while there is a good correspondence between our translated PBH bounds and PBH bounds in literature~\cite{Poulin:2016anj, Lucca:2019rxf, Khan:2025kag, Saha:2024ies}, our constraints in Fig.~\ref{fig:pbh_cons} should be treated as a rough estimate. We leave a full rigorous analysis for evaporating PBHs as well as translation to other motivated DS scenarios for future work.  

\acknowledgments
The authors would like to thank S. Palomares, Hongwan Liu, Dominic Agius for discussions and useful comments on the manuscript.  LLH is
 supported by the Fonds de la Recherche Scientifique F.R.S.-FNRS through a senior research
associate and  acknowledges the support of the FNRS research
grant number J.0134.24. SV is supported by the IISN-FNRS convention « Virgo: physics with gravitational waves ».  All authors are members of BLU-ULB (Brussels Laboratory of the Universe, blu.ulb.be) and acknowledge support of the ARC program of the Federation Wallonie-Bruxelles and  the IISN
convention No. 4.4503.15. Computational resources have been provided by the Consortium
des Équipements de Calcul Intensif (CÉCI), funded by the Fonds de la Recherche Scientifique de Belgique (F.R.S.-FNRS) under Grant No. 2.5020.11 and by the Walloon Region
of Belgium. 

\bibliographystyle{apsrev4-2}
\bibliography{references}

\begin{thebibliography}{47}%
\makeatletter
\providecommand \@ifxundefined [1]{%
 \@ifx{#1\undefined}
}%
\providecommand \@ifnum [1]{%
 \ifnum #1\expandafter \@firstoftwo
 \else \expandafter \@secondoftwo
 \fi
}%
\providecommand \@ifx [1]{%
 \ifx #1\expandafter \@firstoftwo
 \else \expandafter \@secondoftwo
 \fi
}%
\providecommand \natexlab [1]{#1}%
\providecommand \enquote  [1]{``#1''}%
\providecommand \bibnamefont  [1]{#1}%
\providecommand \bibfnamefont [1]{#1}%
\providecommand \citenamefont [1]{#1}%
\providecommand \href@noop [0]{\@secondoftwo}%
\providecommand \href [0]{\begingroup \@sanitize@url \@href}%
\providecommand \@href[1]{\@@startlink{#1}\@@href}%
\providecommand \@@href[1]{\endgroup#1\@@endlink}%
\providecommand \@sanitize@url [0]{\catcode `\\12\catcode `\$12\catcode
  `\&12\catcode `\#12\catcode `\^12\catcode `\_12\catcode `\%12\relax}%
\providecommand \@@startlink[1]{}%
\providecommand \@@endlink[0]{}%
\providecommand \url  [0]{\begingroup\@sanitize@url \@url }%
\providecommand \@url [1]{\endgroup\@href {#1}{\urlprefix }}%
\providecommand \urlprefix  [0]{URL }%
\providecommand \Eprint [0]{\href }%
\providecommand \doibase [0]{https://doi.org/}%
\providecommand \selectlanguage [0]{\@gobble}%
\providecommand \bibinfo  [0]{\@secondoftwo}%
\providecommand \bibfield  [0]{\@secondoftwo}%
\providecommand \translation [1]{[#1]}%
\providecommand \BibitemOpen [0]{}%
\providecommand \bibitemStop [0]{}%
\providecommand \bibitemNoStop [0]{.\EOS\space}%
\providecommand \EOS [0]{\spacefactor3000\relax}%
\providecommand \BibitemShut  [1]{\csname bibitem#1\endcsname}%
\let\auto@bib@innerbib\@empty
\bibitem [{\citenamefont {Shull}\ and\ \citenamefont {van
  Steenberg}(1985)}]{Shull:1985}%
  \BibitemOpen
  \bibfield  {author} {\bibinfo {author} {\bibfnamefont {J.~M.}\ \bibnamefont
  {Shull}}\ and\ \bibinfo {author} {\bibfnamefont {M.~E.}\ \bibnamefont {van
  Steenberg}},\ }\href {https://doi.org/10.1086/163605} {\bibfield  {journal}
  {\bibinfo  {journal} {Astrophys. J.}\ }\textbf {\bibinfo {volume} {298}},\
  \bibinfo {pages} {268} (\bibinfo {year} {1985})}\BibitemShut {NoStop}%
\bibitem [{\citenamefont {Adams}\ \emph {et~al.}(1998)\citenamefont {Adams},
  \citenamefont {Sarkar},\ and\ \citenamefont {Sciama}}]{Adams:1998nr}%
  \BibitemOpen
  \bibfield  {author} {\bibinfo {author} {\bibfnamefont {J.~A.}\ \bibnamefont
  {Adams}}, \bibinfo {author} {\bibfnamefont {S.}~\bibnamefont {Sarkar}},\ and\
  \bibinfo {author} {\bibfnamefont {D.~W.}\ \bibnamefont {Sciama}},\ }\href
  {https://doi.org/10.1046/j.1365-8711.1998.02017.x} {\bibfield  {journal}
  {\bibinfo  {journal} {Mon. Not. Roy. Astron. Soc.}\ }\textbf {\bibinfo
  {volume} {301}},\ \bibinfo {pages} {210} (\bibinfo {year} {1998})},\ \Eprint
  {https://arxiv.org/abs/astro-ph/9805108} {arXiv:astro-ph/9805108}
  \BibitemShut {NoStop}%
\bibitem [{\citenamefont {Chen}\ and\ \citenamefont
  {Kamionkowski}(2004)}]{Chen:2003gz}%
  \BibitemOpen
  \bibfield  {author} {\bibinfo {author} {\bibfnamefont {X.-L.}\ \bibnamefont
  {Chen}}\ and\ \bibinfo {author} {\bibfnamefont {M.}~\bibnamefont
  {Kamionkowski}},\ }\href {https://doi.org/10.1103/PhysRevD.70.043502}
  {\bibfield  {journal} {\bibinfo  {journal} {Phys. Rev.}\ }\textbf {\bibinfo
  {volume} {D70}},\ \bibinfo {pages} {043502} (\bibinfo {year} {2004})},\
  \Eprint {https://arxiv.org/abs/astro-ph/0310473} {arXiv:astro-ph/0310473
  [astro-ph]} \BibitemShut {NoStop}%
\bibitem [{\citenamefont {Padmanabhan}\ and\ \citenamefont
  {Finkbeiner}(2005)}]{Padmanabhan:2005es}%
  \BibitemOpen
  \bibfield  {author} {\bibinfo {author} {\bibfnamefont {N.}~\bibnamefont
  {Padmanabhan}}\ and\ \bibinfo {author} {\bibfnamefont {D.~P.}\ \bibnamefont
  {Finkbeiner}},\ }\href {https://doi.org/10.1103/PhysRevD.72.023508}
  {\bibfield  {journal} {\bibinfo  {journal} {Phys. Rev. D}\ }\textbf {\bibinfo
  {volume} {72}},\ \bibinfo {pages} {023508} (\bibinfo {year} {2005})},\
  \Eprint {https://arxiv.org/abs/astro-ph/0503486} {arXiv:astro-ph/0503486}
  \BibitemShut {NoStop}%
\bibitem [{\citenamefont {Slatyer}\ \emph {et~al.}(2009)\citenamefont
  {Slatyer}, \citenamefont {Padmanabhan},\ and\ \citenamefont
  {Finkbeiner}}]{Slatyer:2009yq}%
  \BibitemOpen
  \bibfield  {author} {\bibinfo {author} {\bibfnamefont {T.~R.}\ \bibnamefont
  {Slatyer}}, \bibinfo {author} {\bibfnamefont {N.}~\bibnamefont
  {Padmanabhan}},\ and\ \bibinfo {author} {\bibfnamefont {D.~P.}\ \bibnamefont
  {Finkbeiner}},\ }\href {https://doi.org/10.1103/PhysRevD.80.043526}
  {\bibfield  {journal} {\bibinfo  {journal} {Phys. Rev. D}\ }\textbf {\bibinfo
  {volume} {80}},\ \bibinfo {pages} {043526} (\bibinfo {year} {2009})},\
  \Eprint {https://arxiv.org/abs/0906.1197} {arXiv:0906.1197 [astro-ph.CO]}
  \BibitemShut {NoStop}%
\bibitem [{\citenamefont {Slatyer}(2016{\natexlab{a}})}]{Slatyer:2015jla}%
  \BibitemOpen
  \bibfield  {author} {\bibinfo {author} {\bibfnamefont {T.~R.}\ \bibnamefont
  {Slatyer}},\ }\href {https://doi.org/10.1103/PhysRevD.93.023527} {\bibfield
  {journal} {\bibinfo  {journal} {Phys. Rev. D}\ }\textbf {\bibinfo {volume}
  {93}},\ \bibinfo {pages} {023527} (\bibinfo {year} {2016}{\natexlab{a}})},\
  \Eprint {https://arxiv.org/abs/1506.03811} {arXiv:1506.03811 [hep-ph]}
  \BibitemShut {NoStop}%
\bibitem [{\citenamefont {Slatyer}(2016{\natexlab{b}})}]{Slatyer:2015kla}%
  \BibitemOpen
  \bibfield  {author} {\bibinfo {author} {\bibfnamefont {T.~R.}\ \bibnamefont
  {Slatyer}},\ }\href {https://doi.org/10.1103/PhysRevD.93.023521} {\bibfield
  {journal} {\bibinfo  {journal} {Phys. Rev. D}\ }\textbf {\bibinfo {volume}
  {93}},\ \bibinfo {pages} {023521} (\bibinfo {year} {2016}{\natexlab{b}})},\
  \Eprint {https://arxiv.org/abs/1506.03812} {arXiv:1506.03812 [astro-ph.CO]}
  \BibitemShut {NoStop}%
\bibitem [{\citenamefont {Slatyer}\ and\ \citenamefont
  {Wu}(2017)}]{Slatyer:2016qyl}%
  \BibitemOpen
  \bibfield  {author} {\bibinfo {author} {\bibfnamefont {T.~R.}\ \bibnamefont
  {Slatyer}}\ and\ \bibinfo {author} {\bibfnamefont {C.-L.}\ \bibnamefont
  {Wu}},\ }\href {https://doi.org/10.1103/PhysRevD.95.023010} {\bibfield
  {journal} {\bibinfo  {journal} {Phys. Rev. D}\ }\textbf {\bibinfo {volume}
  {95}},\ \bibinfo {pages} {023010} (\bibinfo {year} {2017})},\ \Eprint
  {https://arxiv.org/abs/1610.06933} {arXiv:1610.06933 [astro-ph.CO]}
  \BibitemShut {NoStop}%
\bibitem [{\citenamefont {Poulin}\ \emph {et~al.}(2017)\citenamefont {Poulin},
  \citenamefont {Lesgourgues},\ and\ \citenamefont {Serpico}}]{Poulin:2016anj}%
  \BibitemOpen
  \bibfield  {author} {\bibinfo {author} {\bibfnamefont {V.}~\bibnamefont
  {Poulin}}, \bibinfo {author} {\bibfnamefont {J.}~\bibnamefont
  {Lesgourgues}},\ and\ \bibinfo {author} {\bibfnamefont {P.~D.}\ \bibnamefont
  {Serpico}},\ }\href {https://doi.org/10.1088/1475-7516/2017/03/043}
  {\bibfield  {journal} {\bibinfo  {journal} {JCAP}\ }\textbf {\bibinfo
  {volume} {03}},\ \bibinfo {pages} {043}},\ \Eprint
  {https://arxiv.org/abs/1610.10051} {arXiv:1610.10051 [astro-ph.CO]}
  \BibitemShut {NoStop}%
\bibitem [{\citenamefont {Lopez-Honorez}\ \emph {et~al.}(2013)\citenamefont
  {Lopez-Honorez}, \citenamefont {Mena}, \citenamefont {Palomares-Ruiz},\ and\
  \citenamefont {Vincent}}]{Lopez-Honorez:2013cua}%
  \BibitemOpen
  \bibfield  {author} {\bibinfo {author} {\bibfnamefont {L.}~\bibnamefont
  {Lopez-Honorez}}, \bibinfo {author} {\bibfnamefont {O.}~\bibnamefont {Mena}},
  \bibinfo {author} {\bibfnamefont {S.}~\bibnamefont {Palomares-Ruiz}},\ and\
  \bibinfo {author} {\bibfnamefont {A.~C.}\ \bibnamefont {Vincent}},\ }\href
  {https://doi.org/10.1088/1475-7516/2013/07/046} {\bibfield  {journal}
  {\bibinfo  {journal} {JCAP}\ }\textbf {\bibinfo {volume} {07}},\ \bibinfo
  {pages} {046}},\ \Eprint {https://arxiv.org/abs/1303.5094} {arXiv:1303.5094
  [astro-ph.CO]} \BibitemShut {NoStop}%
\bibitem [{\citenamefont {Diamanti}\ \emph {et~al.}(2014)\citenamefont
  {Diamanti}, \citenamefont {Lopez-Honorez}, \citenamefont {Mena},
  \citenamefont {Palomares-Ruiz},\ and\ \citenamefont
  {Vincent}}]{Diamanti:2013bia}%
  \BibitemOpen
  \bibfield  {author} {\bibinfo {author} {\bibfnamefont {R.}~\bibnamefont
  {Diamanti}}, \bibinfo {author} {\bibfnamefont {L.}~\bibnamefont
  {Lopez-Honorez}}, \bibinfo {author} {\bibfnamefont {O.}~\bibnamefont {Mena}},
  \bibinfo {author} {\bibfnamefont {S.}~\bibnamefont {Palomares-Ruiz}},\ and\
  \bibinfo {author} {\bibfnamefont {A.~C.}\ \bibnamefont {Vincent}},\ }\href
  {https://doi.org/10.1088/1475-7516/2014/02/017} {\bibfield  {journal}
  {\bibinfo  {journal} {JCAP}\ }\textbf {\bibinfo {volume} {02}},\ \bibinfo
  {pages} {017}},\ \Eprint {https://arxiv.org/abs/1308.2578} {arXiv:1308.2578
  [astro-ph.CO]} \BibitemShut {NoStop}%
\bibitem [{\citenamefont {Liu}\ \emph {et~al.}(2016)\citenamefont {Liu},
  \citenamefont {Slatyer},\ and\ \citenamefont {Zavala}}]{Liu:2016cnk}%
  \BibitemOpen
  \bibfield  {author} {\bibinfo {author} {\bibfnamefont {H.}~\bibnamefont
  {Liu}}, \bibinfo {author} {\bibfnamefont {T.~R.}\ \bibnamefont {Slatyer}},\
  and\ \bibinfo {author} {\bibfnamefont {J.}~\bibnamefont {Zavala}},\ }\href
  {https://doi.org/10.1103/PhysRevD.94.063507} {\bibfield  {journal} {\bibinfo
  {journal} {Phys. Rev. D}\ }\textbf {\bibinfo {volume} {94}},\ \bibinfo
  {pages} {063507} (\bibinfo {year} {2016})},\ \Eprint
  {https://arxiv.org/abs/1604.02457} {arXiv:1604.02457 [astro-ph.CO]}
  \BibitemShut {NoStop}%
\bibitem [{\citenamefont {Acharya}\ and\ \citenamefont
  {Khatri}(2019)}]{Acharya:2019uba}%
  \BibitemOpen
  \bibfield  {author} {\bibinfo {author} {\bibfnamefont {S.~K.}\ \bibnamefont
  {Acharya}}\ and\ \bibinfo {author} {\bibfnamefont {R.}~\bibnamefont
  {Khatri}},\ }\href {https://doi.org/10.1088/1475-7516/2019/12/046} {\bibfield
   {journal} {\bibinfo  {journal} {JCAP}\ }\textbf {\bibinfo {volume} {12}},\
  \bibinfo {pages} {046}},\ \Eprint {https://arxiv.org/abs/1910.06272}
  {arXiv:1910.06272 [astro-ph.CO]} \BibitemShut {NoStop}%
\bibitem [{\citenamefont {Lucca}\ \emph {et~al.}(2020)\citenamefont {Lucca},
  \citenamefont {Sch{\"o}neberg}, \citenamefont {Hooper}, \citenamefont
  {Lesgourgues},\ and\ \citenamefont {Chluba}}]{Lucca:2019rxf}%
  \BibitemOpen
  \bibfield  {author} {\bibinfo {author} {\bibfnamefont {M.}~\bibnamefont
  {Lucca}}, \bibinfo {author} {\bibfnamefont {N.}~\bibnamefont
  {Sch{\"o}neberg}}, \bibinfo {author} {\bibfnamefont {D.~C.}\ \bibnamefont
  {Hooper}}, \bibinfo {author} {\bibfnamefont {J.}~\bibnamefont
  {Lesgourgues}},\ and\ \bibinfo {author} {\bibfnamefont {J.}~\bibnamefont
  {Chluba}},\ }\href {https://doi.org/10.1088/1475-7516/2020/02/026} {\bibfield
   {journal} {\bibinfo  {journal} {JCAP}\ }\textbf {\bibinfo {volume} {02}},\
  \bibinfo {pages} {026}},\ \Eprint {https://arxiv.org/abs/1910.04619}
  {arXiv:1910.04619 [astro-ph.CO]} \BibitemShut {NoStop}%
\bibitem [{\citenamefont {Bolliet}\ \emph {et~al.}(2021)\citenamefont
  {Bolliet}, \citenamefont {Chluba},\ and\ \citenamefont
  {Battye}}]{Bolliet:2020ofj}%
  \BibitemOpen
  \bibfield  {author} {\bibinfo {author} {\bibfnamefont {B.}~\bibnamefont
  {Bolliet}}, \bibinfo {author} {\bibfnamefont {J.}~\bibnamefont {Chluba}},\
  and\ \bibinfo {author} {\bibfnamefont {R.}~\bibnamefont {Battye}},\ }\href
  {https://doi.org/10.1093/mnras/stab1997} {\bibfield  {journal} {\bibinfo
  {journal} {Mon. Not. Roy. Astron. Soc.}\ }\textbf {\bibinfo {volume} {507}},\
  \bibinfo {pages} {3148} (\bibinfo {year} {2021})},\ \Eprint
  {https://arxiv.org/abs/2012.07292} {arXiv:2012.07292 [astro-ph.CO]}
  \BibitemShut {NoStop}%
\bibitem [{\citenamefont {Capozzi}\ \emph {et~al.}(2023)\citenamefont
  {Capozzi}, \citenamefont {Ferreira}, \citenamefont {Lopez-Honorez},\ and\
  \citenamefont {Mena}}]{Capozzi:2023xie}%
  \BibitemOpen
  \bibfield  {author} {\bibinfo {author} {\bibfnamefont {F.}~\bibnamefont
  {Capozzi}}, \bibinfo {author} {\bibfnamefont {R.~Z.}\ \bibnamefont
  {Ferreira}}, \bibinfo {author} {\bibfnamefont {L.}~\bibnamefont
  {Lopez-Honorez}},\ and\ \bibinfo {author} {\bibfnamefont {O.}~\bibnamefont
  {Mena}},\ }\href {https://doi.org/10.1088/1475-7516/2023/06/060} {\bibfield
  {journal} {\bibinfo  {journal} {JCAP}\ }\textbf {\bibinfo {volume} {06}},\
  \bibinfo {pages} {060}},\ \Eprint {https://arxiv.org/abs/2303.07426}
  {arXiv:2303.07426 [astro-ph.CO]} \BibitemShut {NoStop}%
\bibitem [{\citenamefont {Liu}\ \emph {et~al.}(2023)\citenamefont {Liu},
  \citenamefont {Qin}, \citenamefont {Ridgway},\ and\ \citenamefont
  {Slatyer}}]{Liu:2023nct}%
  \BibitemOpen
  \bibfield  {author} {\bibinfo {author} {\bibfnamefont {H.}~\bibnamefont
  {Liu}}, \bibinfo {author} {\bibfnamefont {W.}~\bibnamefont {Qin}}, \bibinfo
  {author} {\bibfnamefont {G.~W.}\ \bibnamefont {Ridgway}},\ and\ \bibinfo
  {author} {\bibfnamefont {T.~R.}\ \bibnamefont {Slatyer}},\ }\href
  {https://doi.org/10.1103/PhysRevD.108.043531} {\bibfield  {journal} {\bibinfo
   {journal} {Phys. Rev. D}\ }\textbf {\bibinfo {volume} {108}},\ \bibinfo
  {pages} {043531} (\bibinfo {year} {2023})},\ \Eprint
  {https://arxiv.org/abs/2303.07370} {arXiv:2303.07370 [astro-ph.CO]}
  \BibitemShut {NoStop}%
\bibitem [{\citenamefont {Myers}\ \emph {et~al.}(2025)\citenamefont {Myers},
  \citenamefont {Agius}, \citenamefont {Gaggero},\ and\ \citenamefont
  {Ricciardone}}]{Myers:2025pfx}%
  \BibitemOpen
  \bibfield  {author} {\bibinfo {author} {\bibfnamefont {C.}~\bibnamefont
  {Myers}}, \bibinfo {author} {\bibfnamefont {D.}~\bibnamefont {Agius}},
  \bibinfo {author} {\bibfnamefont {D.}~\bibnamefont {Gaggero}},\ and\ \bibinfo
  {author} {\bibfnamefont {A.}~\bibnamefont {Ricciardone}},\ }\href@noop {} {\
  (\bibinfo {year} {2025})},\ \Eprint {https://arxiv.org/abs/2512.10896}
  {arXiv:2512.10896 [astro-ph.CO]} \BibitemShut {NoStop}%
\bibitem [{\citenamefont {Aghanim}\ \emph {et~al.}(2020)\citenamefont {Aghanim}
  \emph {et~al.}}]{Planck:2018vyg}%
  \BibitemOpen
  \bibfield  {author} {\bibinfo {author} {\bibfnamefont {N.}~\bibnamefont
  {Aghanim}} \emph {et~al.} (\bibinfo {collaboration} {Planck}),\ }\href
  {https://doi.org/10.1051/0004-6361/201833910} {\bibfield  {journal} {\bibinfo
   {journal} {Astron. Astrophys.}\ }\textbf {\bibinfo {volume} {641}},\
  \bibinfo {pages} {A6} (\bibinfo {year} {2020})},\ \bibinfo {note} {[Erratum:
  Astron.Astrophys. 652, C4 (2021)]},\ \Eprint
  {https://arxiv.org/abs/1807.06209} {arXiv:1807.06209 [astro-ph.CO]}
  \BibitemShut {NoStop}%
\bibitem [{\citenamefont {Walther}\ \emph {et~al.}(2019)\citenamefont
  {Walther}, \citenamefont {O{\~n}orbe}, \citenamefont {Hennawi},\ and\
  \citenamefont {Luki{\'c}}}]{Walther:2018pnn}%
  \BibitemOpen
  \bibfield  {author} {\bibinfo {author} {\bibfnamefont {M.}~\bibnamefont
  {Walther}}, \bibinfo {author} {\bibfnamefont {J.}~\bibnamefont {O{\~n}orbe}},
  \bibinfo {author} {\bibfnamefont {J.~F.}\ \bibnamefont {Hennawi}},\ and\
  \bibinfo {author} {\bibfnamefont {Z.}~\bibnamefont {Luki{\'c}}},\ }\href
  {https://doi.org/10.3847/1538-4357/aafad1} {\bibfield  {journal} {\bibinfo
  {journal} {Astrophys. J.}\ }\textbf {\bibinfo {volume} {872}},\ \bibinfo
  {pages} {13} (\bibinfo {year} {2019})},\ \Eprint
  {https://arxiv.org/abs/1808.04367} {arXiv:1808.04367 [astro-ph.CO]}
  \BibitemShut {NoStop}%
\bibitem [{\citenamefont {Gaikwad}\ \emph {et~al.}(2020)\citenamefont {Gaikwad}
  \emph {et~al.}}]{Gaikwad:2020art}%
  \BibitemOpen
  \bibfield  {author} {\bibinfo {author} {\bibfnamefont {P.}~\bibnamefont
  {Gaikwad}} \emph {et~al.},\ }\href {https://doi.org/10.1093/mnras/staa907}
  {\bibfield  {journal} {\bibinfo  {journal} {Mon. Not. Roy. Astron. Soc.}\
  }\textbf {\bibinfo {volume} {494}},\ \bibinfo {pages} {5091} (\bibinfo {year}
  {2020})},\ \Eprint {https://arxiv.org/abs/2001.10018} {arXiv:2001.10018
  [astro-ph.CO]} \BibitemShut {NoStop}%
\bibitem [{\citenamefont {Cirelli}\ \emph {et~al.}(2009)\citenamefont
  {Cirelli}, \citenamefont {Iocco},\ and\ \citenamefont
  {Panci}}]{Cirelli:2009bb}%
  \BibitemOpen
  \bibfield  {author} {\bibinfo {author} {\bibfnamefont {M.}~\bibnamefont
  {Cirelli}}, \bibinfo {author} {\bibfnamefont {F.}~\bibnamefont {Iocco}},\
  and\ \bibinfo {author} {\bibfnamefont {P.}~\bibnamefont {Panci}},\ }\href
  {https://doi.org/10.1088/1475-7516/2009/10/009} {\bibfield  {journal}
  {\bibinfo  {journal} {JCAP}\ }\textbf {\bibinfo {volume} {10}},\ \bibinfo
  {pages} {009}},\ \Eprint {https://arxiv.org/abs/0907.0719} {arXiv:0907.0719
  [astro-ph.CO]} \BibitemShut {NoStop}%
\bibitem [{\citenamefont {Liu}\ \emph {et~al.}(2021)\citenamefont {Liu},
  \citenamefont {Qin}, \citenamefont {Ridgway},\ and\ \citenamefont
  {Slatyer}}]{Liu:2020wqz}%
  \BibitemOpen
  \bibfield  {author} {\bibinfo {author} {\bibfnamefont {H.}~\bibnamefont
  {Liu}}, \bibinfo {author} {\bibfnamefont {W.}~\bibnamefont {Qin}}, \bibinfo
  {author} {\bibfnamefont {G.~W.}\ \bibnamefont {Ridgway}},\ and\ \bibinfo
  {author} {\bibfnamefont {T.~R.}\ \bibnamefont {Slatyer}},\ }\href
  {https://doi.org/10.1103/PhysRevD.104.043514} {\bibfield  {journal} {\bibinfo
   {journal} {Phys. Rev. D}\ }\textbf {\bibinfo {volume} {104}},\ \bibinfo
  {pages} {043514} (\bibinfo {year} {2021})},\ \Eprint
  {https://arxiv.org/abs/2008.01084} {arXiv:2008.01084 [astro-ph.CO]}
  \BibitemShut {NoStop}%
\bibitem [{\citenamefont {DeBoer}\ \emph {et~al.}(2017)\citenamefont {DeBoer}
  \emph {et~al.}}]{DeBoer:2016tnn}%
  \BibitemOpen
  \bibfield  {author} {\bibinfo {author} {\bibfnamefont {D.~R.}\ \bibnamefont
  {DeBoer}} \emph {et~al.},\ }\href
  {https://doi.org/10.1088/1538-3873/129/974/045001} {\bibfield  {journal}
  {\bibinfo  {journal} {Publ. Astron. Soc. Pac.}\ }\textbf {\bibinfo {volume}
  {129}},\ \bibinfo {pages} {045001} (\bibinfo {year} {2017})},\ \Eprint
  {https://arxiv.org/abs/1606.07473} {arXiv:1606.07473 [astro-ph.IM]}
  \BibitemShut {NoStop}%
\bibitem [{\citenamefont {Schilizzi}\ \emph {et~al.}(2011)\citenamefont
  {Schilizzi}, \citenamefont {Dewdney},\ and\ \citenamefont
  {Lazio}}]{inproceedings}%
  \BibitemOpen
  \bibfield  {author} {\bibinfo {author} {\bibfnamefont {R.}~\bibnamefont
  {Schilizzi}}, \bibinfo {author} {\bibfnamefont {P.}~\bibnamefont {Dewdney}},\
  and\ \bibinfo {author} {\bibfnamefont {J.}~\bibnamefont {Lazio}}\ }(\bibinfo
  {year} {2011})\ p.\ \bibinfo {pages} {002}\BibitemShut {NoStop}%
\bibitem [{\citenamefont {Facchinetti}\ \emph {et~al.}(2024)\citenamefont
  {Facchinetti}, \citenamefont {Lopez-Honorez}, \citenamefont {Qin},\ and\
  \citenamefont {Mesinger}}]{Facchinetti:2023slb}%
  \BibitemOpen
  \bibfield  {author} {\bibinfo {author} {\bibfnamefont {G.}~\bibnamefont
  {Facchinetti}}, \bibinfo {author} {\bibfnamefont {L.}~\bibnamefont
  {Lopez-Honorez}}, \bibinfo {author} {\bibfnamefont {Y.}~\bibnamefont {Qin}},\
  and\ \bibinfo {author} {\bibfnamefont {A.}~\bibnamefont {Mesinger}},\ }\href
  {https://doi.org/10.1088/1475-7516/2024/01/005} {\bibfield  {journal}
  {\bibinfo  {journal} {JCAP}\ }\textbf {\bibinfo {volume} {01}},\ \bibinfo
  {pages} {005}},\ \Eprint {https://arxiv.org/abs/2308.16656} {arXiv:2308.16656
  [astro-ph.CO]} \BibitemShut {NoStop}%
\bibitem [{\citenamefont {Sun}\ \emph {et~al.}(2025{\natexlab{a}})\citenamefont
  {Sun}, \citenamefont {Foster}, \citenamefont {Liu}, \citenamefont
  {Mu{\~n}oz},\ and\ \citenamefont {Slatyer}}]{Sun:2023acy}%
  \BibitemOpen
  \bibfield  {author} {\bibinfo {author} {\bibfnamefont {Y.}~\bibnamefont
  {Sun}}, \bibinfo {author} {\bibfnamefont {J.~W.}\ \bibnamefont {Foster}},
  \bibinfo {author} {\bibfnamefont {H.}~\bibnamefont {Liu}}, \bibinfo {author}
  {\bibfnamefont {J.~B.}\ \bibnamefont {Mu{\~n}oz}},\ and\ \bibinfo {author}
  {\bibfnamefont {T.~R.}\ \bibnamefont {Slatyer}},\ }\href
  {https://doi.org/10.1103/PhysRevD.111.043015} {\bibfield  {journal} {\bibinfo
   {journal} {Phys. Rev. D}\ }\textbf {\bibinfo {volume} {111}},\ \bibinfo
  {pages} {043015} (\bibinfo {year} {2025}{\natexlab{a}})},\ \Eprint
  {https://arxiv.org/abs/2312.11608} {arXiv:2312.11608 [hep-ph]} \BibitemShut
  {NoStop}%
\bibitem [{\citenamefont {Agius}\ and\ \citenamefont
  {Slatyer}(2025)}]{Agius:2025nfz}%
  \BibitemOpen
  \bibfield  {author} {\bibinfo {author} {\bibfnamefont {D.}~\bibnamefont
  {Agius}}\ and\ \bibinfo {author} {\bibfnamefont {T.~R.}\ \bibnamefont
  {Slatyer}},\ }\href@noop {} {\  (\bibinfo {year} {2025})},\ \Eprint
  {https://arxiv.org/abs/2510.26791} {arXiv:2510.26791 [astro-ph.CO]}
  \BibitemShut {NoStop}%
\bibitem [{\citenamefont {Langhoff}\ \emph {et~al.}(2022)\citenamefont
  {Langhoff}, \citenamefont {Outmezguine},\ and\ \citenamefont
  {Rodd}}]{Langhoff:2022bij}%
  \BibitemOpen
  \bibfield  {author} {\bibinfo {author} {\bibfnamefont {K.}~\bibnamefont
  {Langhoff}}, \bibinfo {author} {\bibfnamefont {N.~J.}\ \bibnamefont
  {Outmezguine}},\ and\ \bibinfo {author} {\bibfnamefont {N.~L.}\ \bibnamefont
  {Rodd}},\ }\href {https://doi.org/10.1103/PhysRevLett.129.241101} {\bibfield
  {journal} {\bibinfo  {journal} {Phys. Rev. Lett.}\ }\textbf {\bibinfo
  {volume} {129}},\ \bibinfo {pages} {241101} (\bibinfo {year} {2022})},\
  \Eprint {https://arxiv.org/abs/2209.06216} {arXiv:2209.06216 [hep-ph]}
  \BibitemShut {NoStop}%
\bibitem [{\citenamefont {Khan}\ \emph {et~al.}(2025)\citenamefont {Khan},
  \citenamefont {Ray}, \citenamefont {Kulkarni},\ and\ \citenamefont
  {Dasgupta}}]{Khan:2025kag}%
  \BibitemOpen
  \bibfield  {author} {\bibinfo {author} {\bibfnamefont {N.~K.}\ \bibnamefont
  {Khan}}, \bibinfo {author} {\bibfnamefont {A.}~\bibnamefont {Ray}}, \bibinfo
  {author} {\bibfnamefont {G.}~\bibnamefont {Kulkarni}},\ and\ \bibinfo
  {author} {\bibfnamefont {B.}~\bibnamefont {Dasgupta}},\ }\href
  {https://doi.org/10.1103/8h9x-hfpk} {\bibfield  {journal} {\bibinfo
  {journal} {Phys. Rev. D}\ }\textbf {\bibinfo {volume} {112}},\ \bibinfo
  {pages} {123019} (\bibinfo {year} {2025})},\ \Eprint
  {https://arxiv.org/abs/2503.15595} {arXiv:2503.15595 [astro-ph.CO]}
  \BibitemShut {NoStop}%
\bibitem [{\citenamefont {Sun}\ \emph {et~al.}(2025{\natexlab{b}})\citenamefont
  {Sun}, \citenamefont {Foster},\ and\ \citenamefont
  {Mu{\~n}oz}}]{Sun:2025ksr}%
  \BibitemOpen
  \bibfield  {author} {\bibinfo {author} {\bibfnamefont {Y.}~\bibnamefont
  {Sun}}, \bibinfo {author} {\bibfnamefont {J.~W.}\ \bibnamefont {Foster}},\
  and\ \bibinfo {author} {\bibfnamefont {J.~B.}\ \bibnamefont {Mu{\~n}oz}},\
  }\href@noop {} {\  (\bibinfo {year} {2025}{\natexlab{b}})},\ \Eprint
  {https://arxiv.org/abs/2509.22772} {arXiv:2509.22772 [hep-ph]} \BibitemShut
  {NoStop}%
\bibitem [{\citenamefont {Saha}\ \emph {et~al.}(2025)\citenamefont {Saha},
  \citenamefont {Singh}, \citenamefont {Parashari},\ and\ \citenamefont
  {Laha}}]{Saha:2024ies}%
  \BibitemOpen
  \bibfield  {author} {\bibinfo {author} {\bibfnamefont {A.~K.}\ \bibnamefont
  {Saha}}, \bibinfo {author} {\bibfnamefont {A.}~\bibnamefont {Singh}},
  \bibinfo {author} {\bibfnamefont {P.}~\bibnamefont {Parashari}},\ and\
  \bibinfo {author} {\bibfnamefont {R.}~\bibnamefont {Laha}},\ }\href
  {https://doi.org/10.1140/epjc/s10052-025-14827-1} {\bibfield  {journal}
  {\bibinfo  {journal} {Eur. Phys. J. C}\ }\textbf {\bibinfo {volume} {85}},\
  \bibinfo {pages} {1117} (\bibinfo {year} {2025})},\ \Eprint
  {https://arxiv.org/abs/2409.10617} {arXiv:2409.10617 [astro-ph.CO]}
  \BibitemShut {NoStop}%
\bibitem [{\citenamefont {Caputo}\ \emph {et~al.}(2025)\citenamefont {Caputo},
  \citenamefont {Park},\ and\ \citenamefont {Yun}}]{Caputo:2025avc}%
  \BibitemOpen
  \bibfield  {author} {\bibinfo {author} {\bibfnamefont {A.}~\bibnamefont
  {Caputo}}, \bibinfo {author} {\bibfnamefont {J.}~\bibnamefont {Park}},\ and\
  \bibinfo {author} {\bibfnamefont {S.}~\bibnamefont {Yun}},\ }\href@noop {} {\
   (\bibinfo {year} {2025})},\ \Eprint {https://arxiv.org/abs/2511.15785}
  {arXiv:2511.15785 [hep-ph]} \BibitemShut {NoStop}%
\bibitem [{\citenamefont {Slatyer}(2013)}]{Slatyer:2012yq}%
  \BibitemOpen
  \bibfield  {author} {\bibinfo {author} {\bibfnamefont {T.~R.}\ \bibnamefont
  {Slatyer}},\ }\href {https://doi.org/10.1103/PhysRevD.87.123513} {\bibfield
  {journal} {\bibinfo  {journal} {Phys. Rev. D}\ }\textbf {\bibinfo {volume}
  {87}},\ \bibinfo {pages} {123513} (\bibinfo {year} {2013})},\ \Eprint
  {https://arxiv.org/abs/1211.0283} {arXiv:1211.0283 [astro-ph.CO]}
  \BibitemShut {NoStop}%
\bibitem [{\citenamefont {Ade}\ \emph {et~al.}(2016)\citenamefont {Ade} \emph
  {et~al.}}]{Planck:2015fie}%
  \BibitemOpen
  \bibfield  {author} {\bibinfo {author} {\bibfnamefont {P.~A.~R.}\
  \bibnamefont {Ade}} \emph {et~al.} (\bibinfo {collaboration} {Planck}),\
  }\href {https://doi.org/10.1051/0004-6361/201525830} {\bibfield  {journal}
  {\bibinfo  {journal} {Astron. Astrophys.}\ }\textbf {\bibinfo {volume}
  {594}},\ \bibinfo {pages} {A13} (\bibinfo {year} {2016})},\ \Eprint
  {https://arxiv.org/abs/1502.01589} {arXiv:1502.01589 [astro-ph.CO]}
  \BibitemShut {NoStop}%
\bibitem [{\citenamefont {Galli}\ \emph {et~al.}(2013)\citenamefont {Galli},
  \citenamefont {Slatyer}, \citenamefont {Valdes},\ and\ \citenamefont
  {Iocco}}]{Galli:2013dna}%
  \BibitemOpen
  \bibfield  {author} {\bibinfo {author} {\bibfnamefont {S.}~\bibnamefont
  {Galli}}, \bibinfo {author} {\bibfnamefont {T.~R.}\ \bibnamefont {Slatyer}},
  \bibinfo {author} {\bibfnamefont {M.}~\bibnamefont {Valdes}},\ and\ \bibinfo
  {author} {\bibfnamefont {F.}~\bibnamefont {Iocco}},\ }\href
  {https://doi.org/10.1103/PhysRevD.88.063502} {\bibfield  {journal} {\bibinfo
  {journal} {Phys. Rev. D}\ }\textbf {\bibinfo {volume} {88}},\ \bibinfo
  {pages} {063502} (\bibinfo {year} {2013})},\ \Eprint
  {https://arxiv.org/abs/1306.0563} {arXiv:1306.0563 [astro-ph.CO]}
  \BibitemShut {NoStop}%
\bibitem [{\citenamefont {Liu}\ \emph {et~al.}(2020)\citenamefont {Liu},
  \citenamefont {Ridgway},\ and\ \citenamefont {Slatyer}}]{Liu:2019bbm}%
  \BibitemOpen
  \bibfield  {author} {\bibinfo {author} {\bibfnamefont {H.}~\bibnamefont
  {Liu}}, \bibinfo {author} {\bibfnamefont {G.~W.}\ \bibnamefont {Ridgway}},\
  and\ \bibinfo {author} {\bibfnamefont {T.~R.}\ \bibnamefont {Slatyer}},\
  }\href {https://doi.org/10.1103/PhysRevD.101.023530} {\bibfield  {journal}
  {\bibinfo  {journal} {Phys. Rev. D}\ }\textbf {\bibinfo {volume} {101}},\
  \bibinfo {pages} {023530} (\bibinfo {year} {2020})},\ \Eprint
  {https://arxiv.org/abs/1904.09296} {arXiv:1904.09296 [astro-ph.CO]}
  \BibitemShut {NoStop}%
\bibitem [{\citenamefont {Slatyer}(2022)}]{Slatyer22}%
  \BibitemOpen
  \bibfield  {author} {\bibinfo {author} {\bibfnamefont {T.~R.}\ \bibnamefont
  {Slatyer}},\ }\href {https://agenda.irmp.ucl.ac.be/event/4622/timetable/}
  {\bibinfo {title} {{Georges Lemaître Chair lecture series}}},\ \bibinfo
  {howpublished} {Lecture notes} (\bibinfo {year} {2022})\BibitemShut {NoStop}%
\bibitem [{\citenamefont {Lopez-Honorez}(2025)}]{GGI:2025Lectures}%
  \BibitemOpen
  \bibfield  {author} {\bibinfo {author} {\bibfnamefont {L.}~\bibnamefont
  {Lopez-Honorez}},\ }\href {https://physth.ulb.be/llh/GGI25_3.pdf} {\bibinfo
  {title} {{GGI School on the Theory of Fundamental Interactions 2025}}},\
  \bibinfo {howpublished} {Lecture notes} (\bibinfo {year} {2025})\BibitemShut
  {NoStop}%
\bibitem [{\citenamefont {Palomares~Ruiz}(2026)}]{Palomares26}%
  \BibitemOpen
  \bibfield  {author} {\bibinfo {author} {\bibfnamefont {S.}~\bibnamefont
  {Palomares~Ruiz}},\ }\href
  {https://indico.global/event/15506/contributions/143655/} {\bibinfo {title}
  {{Exotic energy injection during cosmic dawn }}},\ \bibinfo {howpublished}
  {talk at COSNP 2026} (\bibinfo {year} {2026})\BibitemShut {NoStop}%
\bibitem [{\citenamefont {Sun}\ and\ \citenamefont
  {Slatyer}(2023)}]{Sun:2022djj}%
  \BibitemOpen
  \bibfield  {author} {\bibinfo {author} {\bibfnamefont {Y.}~\bibnamefont
  {Sun}}\ and\ \bibinfo {author} {\bibfnamefont {T.~R.}\ \bibnamefont
  {Slatyer}},\ }\href {https://doi.org/10.1103/PhysRevD.107.063541} {\bibfield
  {journal} {\bibinfo  {journal} {Phys. Rev. D}\ }\textbf {\bibinfo {volume}
  {107}},\ \bibinfo {pages} {063541} (\bibinfo {year} {2023})},\ \Eprint
  {https://arxiv.org/abs/2207.06425} {arXiv:2207.06425 [hep-ph]} \BibitemShut
  {NoStop}%
\bibitem [{\citenamefont {Abdurashidova}\ \emph {et~al.}(2022)\citenamefont
  {Abdurashidova} \emph {et~al.}}]{HERA:2021bsv}%
  \BibitemOpen
  \bibfield  {author} {\bibinfo {author} {\bibfnamefont {Z.}~\bibnamefont
  {Abdurashidova}} \emph {et~al.} (\bibinfo {collaboration} {HERA}),\ }\href
  {https://doi.org/10.3847/1538-4357/ac1c78} {\bibfield  {journal} {\bibinfo
  {journal} {Astrophys. J.}\ }\textbf {\bibinfo {volume} {925}},\ \bibinfo
  {pages} {221} (\bibinfo {year} {2022})},\ \Eprint
  {https://arxiv.org/abs/2108.02263} {arXiv:2108.02263 [astro-ph.CO]}
  \BibitemShut {NoStop}%
\bibitem [{\citenamefont {Cima}\ and\ \citenamefont
  {D'Eramo}(2025)}]{Cima:2025zmc}%
  \BibitemOpen
  \bibfield  {author} {\bibinfo {author} {\bibfnamefont {F.}~\bibnamefont
  {Cima}}\ and\ \bibinfo {author} {\bibfnamefont {F.}~\bibnamefont {D'Eramo}},\
  }\href@noop {} {\  (\bibinfo {year} {2025})},\ \Eprint
  {https://arxiv.org/abs/2507.10664} {arXiv:2507.10664 [hep-ph]} \BibitemShut
  {NoStop}%
\bibitem [{\citenamefont {Agius}\ \emph {et~al.}(2025)\citenamefont {Agius},
  \citenamefont {Essig}, \citenamefont {Gaggero}, \citenamefont
  {Palomares-Ruiz}, \citenamefont {Suczewski},\ and\ \citenamefont
  {Valli}}]{Agius:2025xbj}%
  \BibitemOpen
  \bibfield  {author} {\bibinfo {author} {\bibfnamefont {D.}~\bibnamefont
  {Agius}}, \bibinfo {author} {\bibfnamefont {R.}~\bibnamefont {Essig}},
  \bibinfo {author} {\bibfnamefont {D.}~\bibnamefont {Gaggero}}, \bibinfo
  {author} {\bibfnamefont {S.}~\bibnamefont {Palomares-Ruiz}}, \bibinfo
  {author} {\bibfnamefont {G.}~\bibnamefont {Suczewski}},\ and\ \bibinfo
  {author} {\bibfnamefont {M.}~\bibnamefont {Valli}},\ }\href@noop {} {\
  (\bibinfo {year} {2025})},\ \Eprint {https://arxiv.org/abs/2510.14877}
  {arXiv:2510.14877 [astro-ph.CO]} \BibitemShut {NoStop}%
\bibitem [{\citenamefont {Poulin}\ \emph {et~al.}(2015)\citenamefont {Poulin},
  \citenamefont {Serpico},\ and\ \citenamefont {Lesgourgues}}]{Poulin:2015pna}%
  \BibitemOpen
  \bibfield  {author} {\bibinfo {author} {\bibfnamefont {V.}~\bibnamefont
  {Poulin}}, \bibinfo {author} {\bibfnamefont {P.~D.}\ \bibnamefont
  {Serpico}},\ and\ \bibinfo {author} {\bibfnamefont {J.}~\bibnamefont
  {Lesgourgues}},\ }\href {https://doi.org/10.1088/1475-7516/2015/12/041}
  {\bibfield  {journal} {\bibinfo  {journal} {JCAP}\ }\textbf {\bibinfo
  {volume} {12}},\ \bibinfo {pages} {041}},\ \Eprint
  {https://arxiv.org/abs/1508.01370} {arXiv:1508.01370 [astro-ph.CO]}
  \BibitemShut {NoStop}%
\bibitem [{\citenamefont {Lopez-Honorez}\ \emph {et~al.}(2016)\citenamefont
  {Lopez-Honorez}, \citenamefont {Mena}, \citenamefont {Molin{\'e}},
  \citenamefont {Palomares-Ruiz},\ and\ \citenamefont
  {Vincent}}]{Lopez-Honorez:2016sur}%
  \BibitemOpen
  \bibfield  {author} {\bibinfo {author} {\bibfnamefont {L.}~\bibnamefont
  {Lopez-Honorez}}, \bibinfo {author} {\bibfnamefont {O.}~\bibnamefont {Mena}},
  \bibinfo {author} {\bibfnamefont {{\'A}.}~\bibnamefont {Molin{\'e}}},
  \bibinfo {author} {\bibfnamefont {S.}~\bibnamefont {Palomares-Ruiz}},\ and\
  \bibinfo {author} {\bibfnamefont {A.~C.}\ \bibnamefont {Vincent}},\ }\href
  {https://doi.org/10.1088/1475-7516/2016/08/004} {\bibfield  {journal}
  {\bibinfo  {journal} {JCAP}\ }\textbf {\bibinfo {volume} {08}},\ \bibinfo
  {pages} {004}},\ \Eprint {https://arxiv.org/abs/1603.06795} {arXiv:1603.06795
  [astro-ph.CO]} \BibitemShut {NoStop}%
\bibitem [{\citenamefont {Carr}\ \emph {et~al.}(2010)\citenamefont {Carr},
  \citenamefont {Kohri}, \citenamefont {Sendouda},\ and\ \citenamefont
  {Yokoyama}}]{Carr:2009jm}%
  \BibitemOpen
  \bibfield  {author} {\bibinfo {author} {\bibfnamefont {B.~J.}\ \bibnamefont
  {Carr}}, \bibinfo {author} {\bibfnamefont {K.}~\bibnamefont {Kohri}},
  \bibinfo {author} {\bibfnamefont {Y.}~\bibnamefont {Sendouda}},\ and\
  \bibinfo {author} {\bibfnamefont {J.}~\bibnamefont {Yokoyama}},\ }\href
  {https://doi.org/10.1103/PhysRevD.81.104019} {\bibfield  {journal} {\bibinfo
  {journal} {Phys. Rev. D}\ }\textbf {\bibinfo {volume} {81}},\ \bibinfo
  {pages} {104019} (\bibinfo {year} {2010})},\ \Eprint
  {https://arxiv.org/abs/0912.5297} {arXiv:0912.5297 [astro-ph.CO]}
  \BibitemShut {NoStop}%
\end{thebibliography}%

\end{document}